\documentclass{aa}  
\usepackage{ifthen}     
\usepackage{graphicx}
\usepackage{natbib}
\usepackage{latexsym}

\bibpunct{(}{)}{;}{a}{}{,}

\newcommand{\ie} {{\em i.e.}}
\newcommand{\eg} {{e.g.}}

\newcommand{\emm}[1]{\ensuremath{#1}}   
\newcommand{\emr}[1]{\emm{\mathrm{#1}}} 

\newcommand{\nds}[1]{\emm{\displaystyle#1}} 

\newcommand{\paren}[1]  {\nds{\left(  #1 \right) }} 
\newcommand{\cbrace}[1] {\nds{\left\{ #1 \right\}}} 
\newcommand{\bracket}[1]{\nds{\left[  #1 \right] }} 
\newcommand{\sbracket}[1]{\emm{\left[  #1 \right]}} 

\newcommand{\Definition}{\emm{\equiv}} 
\newcommand{\about}{\emm{\sim}}        

\newcommand{\df}[1]{\emm{\emr{d}#1}}         
\newcommand{\Diff}[1]{\emm{\partial #1}} 

\newcommand{\sint}[2]{\int_{#1}#2\,\df{#1}}
\newcommand{\dint}[4][\!\!\!\!\!]{\int#1\int_{#2#3}\!\!\!\!#4\,\df{#2}\df{#3}}

\newcommand{\CONV}[3]{\emm{\cbrace{#1\star#2}
    \ifthenelse{\equal{#3}{}}{}{\!\paren{#3}}}}
\newcommand{\conv}[4]{\sint{#3}{#1(#4-#3)\,#2(#3)}}

\newcommand{\expi}[2][-]{\emm{\emr{e}^{#1i2\pi#2}}} 
\newcommand{\dexp}[1]{\expi[-]{#1}} 
\newcommand{\iexp}[1]{\expi[+]{#1}} 
\newcommand{\DFTint}[3]{\sint{#1}{#2\,\dexp{#1#3}}} 
\newcommand{\IFTint}[3]{\sint{#1}{#2\,\iexp{#1#3}}} 

\newcommand{\FTop}[4]{\ifthenelse{\equal{#3}{}}{\emm{#1\supset#2}}
  {\emm{#1\,\,\mbox{\raisebox{-1.1ex}{\emm{\stackrel{\scriptstyle#3}{\stackrel{\textstyle\supset}{\scriptstyle#4}}}}}\,\,#2}}}

\newcommand{\dnnftop}[1]{\emm{\overline{#1}}} 
\newcommand{\dftnnop}[1]{\emm{\underline{#1}}} 

\newcommand{\iftnnop}[1]{\raisebox{-1.1ex}{\emm{\stackrel{\nds{#1}}{\scriptstyle\sim}}}} 

\newcommand{\updown}[2]{\ifthenelse{\equal{#1}{}}{}{_{#1}}\ifthenelse{\equal{#2}{}}{}{^{#2}}}
\newcommand{\meas}[3]{\emm{#1\updown{#2}{#3}}}

\newcommand{\dftmeas}[3]{\emm{\dnnftop{#1}\updown{#2}{#3}}}

\newcommand{\dnnftmeas}[3]{\emm{\dnnftop{#1}\updown{#2}{#3}}}
\newcommand{\dftnnmeas}[3]{\emm{\dftnnop{#1}\updown{#2}{#3}}}
\newcommand{\dftftmeas}[3]{\emm{\dftnnop{\dnnftop{#1}}\updown{#2}{#3}}}

\newcommand{\iftnnmeas}[3]{\emm{\iftnnop{#1}\updown{#2}{#3}}}

\newcommand{\dirac}[1]{\emm{\delta{#1}}}
\newcommand{\boxcar}{\emm{\Pi}}
\newcommand{\sinc}{\emr{sinc}}
\newcommand{\abs}[1]{\emm{\left| #1 \right|}} 
\newcommand{\average}[2][]{\emm{\left\langle #2 \right\rangle\ifthenelse{\equal{#1}{}}{}{_{#1}}\!}} 
\newcommand{\linepath}{\dirac{\bracket{\uf-(\up+\us)}}}

\newcommand{\as}{\emm{\alpha_s}}     
\newcommand{\ass}{\emm{\alpha'_s}}   
\newcommand{\asss}{\emm{\alpha''_s}} 
\newcommand{\ap}{\emm{\alpha_p}}     
\newcommand{\app}{\emm{\alpha'_p}}   
\newcommand{\af}{\emm{\alpha}}       
\newcommand{\aff}{\emm{\alpha'}}     
\newcommand{\afff}{\emm{\alpha''}}   
\newcommand{\ai}{\emm{\alpha_i}}     
\newcommand{\ak}{\emm{\alpha_k}}     
\newcommand{\ab}{\emm{\beta}}        

\newcommand{\das}{\Diff{\as}}
\newcommand{\dass}{\Diff{\ass}}
\newcommand{\dasss}{\Diff{\asss}}
\newcommand{\dap}{\Diff{\ap}}
\newcommand{\daf}{\Diff{\af}}
\newcommand{\daff}{\Diff{\aff}}
\newcommand{\dafff}{\Diff{\afff}}

\newcommand{\us}{\emm{u_s}}     
\newcommand{\uss}{\emm{u'_s}}   
\newcommand{\usss}{\emm{u''_s}} 
\newcommand{\up}{\emm{u_p}}     
\newcommand{\upp}{\emm{u'_p}}   
\newcommand{\uf}{\emm{u}}       
\newcommand{\uff}{\emm{u'}}     
\newcommand{\ufff}{\emm{u''}}   

\newcommand{\dus}{\Diff{\us}}
\newcommand{\duss}{\Diff{\uss}}
\newcommand{\dusss}{\Diff{\usss}}
\newcommand{\dup}{\Diff{\up}}
\newcommand{\duf}{\Diff{\uf}}
\newcommand{\duff}{\Diff{\uff}}
\newcommand{\dufff}{\Diff{\ufff}}

\newcommand{\Bg}[2][]{\meas{\mathcal{B}}{#2}{#1}} 
\newcommand{\Ig}[2][]{\meas{\mathcal{I}}{#2}{#1}} 

\newcommand{\I}[2][]{\meas{I}{#2}{#1}}             
\newcommand{\Ift}[2][]{\dftmeas{I}{#2}{#1}}        

\newcommand{\B}[2][]{\meas{B}{#2}{#1}}             
\newcommand{\Bft}[2][]{\dftmeas{B}{#2}{#1}}        

\newcommand{\Bv}[2][]{\meas{b}{#2}{#1}}            
\newcommand{\Bvft}[2][]{\dftmeas{b}{#2}{#1}}       

\newcommand{\V}[2][]{\meas{V}{#2}{#1}}             
\newcommand{\Vft}[2][]{\dftmeas{V}{#2}{#1}}        
\newcommand{\Vftnn}[2][]{\iftnnmeas{V}{#2}{#1}}    

\renewcommand{\S}[2][]{\meas{S}{#2}{#1}}               
\newcommand{\Sft}[2][]{\dftmeas{S}{#2}{#1}}            

\newcommand{\SV}[2][]{\meas{SV}{#2}{#1}}               
\newcommand{\SVft}[2][]{\dftmeas{SV}{#2}{#1}}          

\newcommand{\SB}[2][]{\meas{\Sigma}{#2}{#1}}           
\newcommand{\SBnnft}[2][]{\dnnftmeas{\Sigma}{#2}{#1}}  
\newcommand{\SBftft}[2][]{\dftftmeas{\Sigma}{#2}{#1}}  

\newcommand{\Wa}[2][]{\meas{\Omega}{#2}{#1}}          
\newcommand{\Waftnn}[2][]{\dftnnmeas{\Omega}{#2}{#1}} 
\newcommand{\Waftft}[2][]{\dftftmeas{\Omega}{#2}{#1}} 
\newcommand{\Waa}[2][]{\meas{\omega}{#2}{#1}}         
\newcommand{\Waaft}[2][]{\dftmeas{\omega}{#2}{#1}}    

\newcommand{\Wu}[2][]{\meas{W}{#2}{#1}}               
\newcommand{\wu}[2][]{\meas{w}{#2}{#1}}               %

\newcommand{\Wm}[1][]{\meas{\Omega}{\emr{mos}}{#1}}   
\newcommand{\Wmu}[1][]{\meas{W}{\emr{mos}}{#1}}       
\newcommand{\wm}[2][]{\meas{w}{#2}{#1}}          %

\newcommand{\Isf}[1][]{\meas{I}{\emr{sfd}}{#1}}        
\newcommand{\Isfft}[1][]{\dftmeas{I}{\emr{sfd}}{#1}}   

\newcommand{\Iwf}[1][]{\meas{I}{\emr{dirty}}{#1}}      
\newcommand{\Iwfft}[1][]{\dftmeas{I}{\emr{dirty}}{#1}} 

\newcommand{\Ds}[2][]{\meas{\Delta}{#2}{#1}}      

\newcommand{\Dw}[2][]{\meas{D}{#2}{#1}}           
\newcommand{\Dwft}[2][]{\dftmeas{D}{#2}{#1}}      
\newcommand{\Dwftnn}[2][]{\dftnnmeas{D}{#2}{#1}}  
\newcommand{\Dwnnft}[2][]{\dnnftmeas{D}{#2}{#1}}  
\newcommand{\Dwftft}[2][]{\dftftmeas{D}{#2}{#1}}  

\newcommand{\G}[2][]{\meas{\mathcal{G}}{#2}{#1}}  

\newcommand{\gu}[2][]{\meas{g}{#2}{#1}}           

\renewcommand{\ga}[2][]{\meas{\gamma}{#2}{#1}}    
\newcommand{\gaft}[2][]{\dftmeas{\ga{}}{#2}{#1}}  

\newcommand{\fb}{\emm{f_\emr{b}}}

\newcommand{\nsamp}{\emm{n_\emr{samp}}}

\newcommand{\dmin}{\emm{d_\emr{min}}}
\newcommand{\dmax}{\emm{d_\emr{max}}}
\newcommand{\Asynth}{\emm{\theta_\emr{syn}}}

\newcommand{\dprim}{\emm{d_\emr{prim}}}
\newcommand{\Aprim}{\emm{\theta_\emr{prim}}}
\newcommand{\Afwhm}{\emm{\theta_\emr{fwhm}}}

\newcommand{\dalias}{\emm{d_\emr{alias}}}
\newcommand{\Aalias}{\emm{\theta_\emr{alias}}}

\newcommand{\dfield}{\emm{d_\emr{field}}}
\newcommand{\Afield}{\emm{\theta_\emr{field}}}

\newcommand{\dimage}{\emm{d_\emr{image}}}
\newcommand{\Aimage}{\emm{\theta_\emr{image}}}

\renewcommand{\Im}[1][]{\meas{I}{\emr{mos}}{#1}}  
\newcommand{\Nm}[1][]{\meas{N}{\emr{mos}}{#1}}    
\newcommand{\Rm}[2][]{\meas{R}{#2}{#1}}           
\newcommand{\SNRm}[2][]{\meas{\emr{SNR}}{#2}{#1}} 
\newcommand{\Nsf}[2][]{\meas{\sigma}{#2}{#1}}     
\newcommand{\Dm}[1][]{\meas{D}{\emr{mos}}{#1}}    
\newcommand{\Dmftft}[1]{\dftftmeas{D}{\emr{mos}}{#1}}    %

\newcommand{\Vsd}[1]{\meas{V}{\emr{sd}}{#1}}             %
\newcommand{\Wsd}[1]{\meas{W}{\emr{sd}}{#1}}             %

\newcommand{\Isd}[1]{\meas{I}{\emr{sd}}{#1}}             %
\newcommand{\Isdft}[1]{\dftmeas{I}{\emr{sd}}{#1}}        %

\newcommand{\Bsd}[1]{\meas{B}{\emr{sd}}{#1}}             %
\newcommand{\Bsdft}[1]{\dftmeas{B}{\emr{sd}}{#1}}        %

\newcommand{\Ssd}[1]{\meas{S}{\emr{sd}}{#1}}             %
\newcommand{\Ssdft}[1]{\dftmeas{S}{\emr{sd}}{#1}}        %

\newcommand{\Dsdftft}[1]{\dftftmeas{D}{\emr{sd}}{#1}}    %

\newcommand{\Whyb}[1]{\meas{W}{\emr{hyb}}{#1}}           %

\newcommand{\Shyb}[1]{\meas{S}{\emr{hyb}}{#1}}           %
\newcommand{\Shybft}[1]{\dftmeas{S}{\emr{hyb}}{#1}}      %

\newcommand{\Ihyb}[1]{\meas{I}{\emr{hyb}}{#1}}           %
\newcommand{\Ihybft}[1]{\dftmeas{I}{\emr{hyb}}{#1}}      %

\newcommand{\Bhyb}[1]{\meas{B}{\emr{hyb}}{#1}}           %
\newcommand{\Bhybft}[1]{\dftmeas{B}{\emr{hyb}}{#1}}      %

\newcommand{\Dhyb}[1]{\meas{D}{\emr{hyb}}{#1}}           %
\newcommand{\Dhybftft}[1]{\dftftmeas{D}{\emr{hyb}}{#1}}  %

\newcommand{\A}[2][]{\meas{A}{#2}{#1}}                 
\newcommand{\Aft}[2][]{\dftmeas{A}{#2}{#1}}            

\newcommand{\Beff}[1][]{\meas{B}{\emr{eff}}{#1}}       
\newcommand{\Beffft}[1][]{\dftmeas{B}{\emr{eff}}{#1}}  

\newcommand{\wearth}{\emm{\omega_\emr{earth}}}
\newcommand{\dt}{\emm{\delta t}}
\newcommand{\hdt}{\emm{\dt/2}}
\newcommand{\tzero}{\emm{t_0}}
\newcommand{\tone}{\emm{\tzero-\hdt}}
\newcommand{\ttwo}{\emm{\tzero+\hdt}}
\newcommand{\Tave}[3]{\frac{1}{\dt}\int_{#1}^{#2}#3\,\df{t}}
\newcommand{\asave}{\emm{\hat{\alpha}_s}}      
\newcommand{\upave}{\emm{\hat{u}_p}}           
\newcommand{\vslew}{\emm{v_\emr{slew}}} 
\newcommand{\daslew}{\emm{\delta \as}} 

\renewcommand{\wp}{\emm{w}}             
\newcommand{\Pw}[2][]{\meas{P}{#2}{#1}} 
\newcommand{\Acenter}{\emm{\theta_\emr{center}}}
\newcommand{\wavelength}{\emm{\lambda}}

\newcommand{\TabScales}{%
  \begin{table}
    \centering
    \caption{Definition of the $uv$ and sky scales relevant to wide-field 
      interferometric imaging.}
    \begin{tabular}{cc}
      \hline %
      \hline %
      Symbol & Definition \\
      $[\wavelength$,rad$]$\tablefootmark{a} & Conjugate $uv$ and angular scale \\ 
      \hline %
      $\dmax,\Asynth$   & Maximum baseline length \& Synthesized beam\\
      $\dprim,\Aprim$   & Antenna diameter \& Primary beamwidth \\
      $\dalias,\Aalias$ & Minimum image size for tolerable aliasing\\
      $\dfield,\Afield$ & Targeted field of view \\
      $\dimage,\Aimage$ & Final image size \\
      \hline %
    \end{tabular}
    \tablefoot{\tablefoottext{a}{The chosen units (radians for $\theta$ and wavelength for $d$)
        imply that the conjugate scales are linked through $\theta = 1/d$, 
        instead of the usual $\theta = \wavelength/d$.}}
    \label{tab:scales}
  \end{table}}

\newcommand{\TabSymbolWideField}{%
  \begin{table}
    \centering
    \caption{Definition of the symbols used to expose the wide-field synthesis
      formalism.}
    \begin{tabular}{clc}
      \hline %
      \hline %
      \multicolumn{2}{l}{Symbol \& Definition} & Plane(s)\tablefootmark{a} \\ 
      \hline %
      \as{} & Scanned angle                       & sky  \\
      \us{} & Scanned spatial frequency           & $uv$ \\
      \ap{} & Phased angle                        & sky  \\
      \up{} & Phased spatial frequency            & $uv$ \\
      \I{}  & Sky brightness                      & sky \\
      \B{}  & Primary beam                        & sky \\
      \V{}  & Visibility function                 & $uv$ \& sky \\
      \S{}  & Sampling function                   & $uv$ \& sky \\
      \Ds{} & Set of single-field dirty beams     & sky \& sky \\
      \Dw{} & Set of wide-field dirty beams       & sky \& sky \\
      \Wa{} & Sky-plane weighting function        & sky \& sky \\
      \Wu{} & $uv$-plane weighting function $(\FTop{\Wa{}}{\Wu{}}{}{})$ & $uv$ \& $uv$ \\
      \G{}  & Gridding function $(=\gu{}\,\ga{})$ & $uv$ \& sky \\
      \gu{} & $uv$-plane gridding function        & $uv$ \\
      \ga{} & Sky-plane gridding function         & sky \\
      \Iwf{} & Wide-field dirty image             & sky \\
      \hline %
    \end{tabular}
    \tablefoot{\tablefoottext{a}{Planes of definition of the associated symbols.}}
    \label{tab:symbols:widefield}
  \end{table}}

\newcommand{\TabSymbolMosaicking}{%
  \begin{table}
    \centering
    \caption{Definition of the symbols used to expose the mosaicking
      formalism.}
    \begin{tabular}{clc}
      \hline %
      \hline %
      \multicolumn{2}{l}{Symbol \& Definition} & Plane(s) \\ 
      \hline %
      \as{} & Scanned angle                    & sky  \\
      \us{} & Scanned spatial frequency        & $uv$ \\
      \ap{} & Phased angle                     & sky  \\
      \up{} & Phased spatial frequency         & $uv$ \\
      \I{}  & Sky brightness                   & sky \\
      \B{}  & Primary beam                     & sky \\
      \V{}  & Visibility function              & $uv$ \& sky \\
      \S{}  & Sampling function                & $uv$ \& sky \\
      \Ds{} & Set of single-field dirty beams  & sky \& sky \\
      \hline %
      \Isf{}& Set of single-field dirty images   & sky \& sky \\
      \Dm{} & Set of mosaicked dirty beams       & sky \& sky \\
      \Wm{} & Mosaicking sky weighting function  & sky \& sky \\
      \Wmu{}& Mosaicking $uv$ weighting function & $uv$ \& $uv$ \\
      \Im{} & Mosaicked dirty image              & sky \\
      \Nm{} & Mosaicked noise image              & sky \\
      \Rm{} & Residual image                     & sky \\
      \SNRm{} & Signal-to-noise ratio image      & sky \\
      \hline %
    \end{tabular}
    \tablefoot{This table uses similar conventions as 
      Table~\ref{tab:symbols:widefield}. The top part repeats
      the symbols which have the same meaning in both the mosaicking and the
      wide-field synthesis formalisms, while the bottom part defines symbols
      that either have a different meaning in both formalisms or are used
      only in the mosaicking formalism.}
    \label{tab:symbols:mosaicking}
  \end{table}}

\newcommand{\TabAliasing}{%
  \begin{table}
    \centering
    \caption{Minimum sizes of the dirty beam images to get an image
      fidelity or a dynamic range greater than a given value.}
    \begin{tabular}{crrr}
      \hline %
      \hline %
      Minimum fidelity & \multicolumn{2}{c}{$\Aalias/\Afwhm$\tablefootmark{a}} \\
      or dynamic range & $(\fb=0)$\tablefootmark{b} & $(\fb=0.0625)$ & $(\fb=0.1)$\\
      \hline %
      $10^{2}$ &  2.2 &    2.2  &    2.2\\
      $10^{3}$ &  3.5 &    3.7  &    6.6\\
      $10^{4}$ &  8.4 &   13.4  &   13.7\\
      $10^{5}$ & 19.8 & $>20.0$ & $>20.0$\\
      \hline %
    \end{tabular}
    \tablefoot{\tablefoottext{a}{The image sizes are expressed in units of the primary beam full
        width at half maxium.} \tablefoottext{b}{The computation is done for 3 different
        ratios of the secondary-to-primary diameters (\ie{} \fb{}, the antenna
        blockage factors). The values are derived from the modeling of the antenna
        power patterns shown in Fig.~\ref{fig:primbeam}.}}
    \label{tab:aliasing}
  \end{table}}

\newcommand{\TabSampling}{%
  \begin{table}
    \centering
    \caption{Interval ranges of definition and associated sampling rates 
      for the used functions.}
    \begin{tabular}{lll}
      \hline %
      \hline %
      Functions & Intervals & Samplings \\
      \hline %
      Visibilities & $\abs{\up}   \le \dmax$     & $\dup   = 2\,\dalias/\nsamp$ \\
                   & $\abs{\ap}   \le \Aalias/2$ & $\dap   = \Asynth/\nsamp$ \\[\smallskipamount]
                   & $\abs{\us}   \le \dprim$    & $\dus   = 2\,\dimage/\nsamp$ \\
                   & $\abs{\as}   \le \Aimage/2$ & $\das   = \Aprim/\nsamp$  \\[\smallskipamount]
      Primary beam & $\abs{\uss}  \le \dprim$    & $\duss  = 2\,\dalias/\nsamp$ \\
                   & $\abs{\ass}  \le \Aalias/2$ & $\dass  = \Aprim/\nsamp$  \\[\smallskipamount]
                   & $\abs{\usss} \le \dprim$    & $\dusss = 2\,\dalias/\nsamp$ \\
                   & $\abs{\asss} \le \Aalias/2$ & $\dasss = \Aprim/\nsamp$  \\[\smallskipamount]
      Dirty image  & $\abs{\uf}   \le \dmax$     & $\duf   = 2\,\dimage/\nsamp$ \\
                   & $\abs{\af}   \le \Aimage/2$ & $\daf   = \Asynth/\nsamp$ \\[\smallskipamount]
      Dirty beam   & $\abs{\uff}  \le \dmax$     & $\duff  = 2\,\dimage/\nsamp$ \\
                   & $\abs{\aff}  \le \Aimage/2$ & $\daff  = \Asynth/\nsamp$ \\[\smallskipamount]
                   & $\abs{\ufff} \le \dprim$    & $\dufff = 2\,\dimage/\nsamp$ \\
                   & $\abs{\afff} \le \Aimage/2$ & $\dafff = \Aprim/\nsamp$  \\[\smallskipamount]
      \hline %
    \end{tabular}
    \label{tab:samplings}
  \end{table}}

\newcommand{\TabSymbolShortSpacings}{%
  \begin{table}
    \centering
    \caption{Definition of the symbols used to expose the processing of the
      short spacings.}
    \begin{tabular}{llc}
      \hline %
      \hline %
      \multicolumn{2}{l}{Symbol \& Definition} & Plane(s) \\ 
      \hline %
      \Isd{}  & Measured single-dish intensity            & sky \\
      \Bsd{}  & Single-dish antenna power pattern         & sky \\
      \Ssd{}  & Single-dish sampling function             & sky \\
      \Vsd{}  & Single-dish visibility function           & sky \\
      \Wsd{}  & Single-dish $uv$-plane weighting function & $uv$ \\
      \hline %
      \Bv{i}  & Voltage pattern of antenna $i$                                  & sky \\
      \B{ij}  & Power pattern of antenna $i$ and $j$ $(=\Bv{i}\,\Bv[\star]{j})$ & sky \\
      \V{ij}  & Visibility between antenna $i$ and $j$                          & $uv$ \& sky \\
      \hline %
      \Ihyb{} & Hybrid dirty image                   & sky \\
      \Bhyb{} & Hybrid antenna power pattern         & sky \\
      \Shyb{} & Hybrid sampling function             & $uv$ \& sky \\
      \Whyb{} & Hybrid $uv$-plane weighting function & $uv$ \& $uv$ \\
      \Dhyb{} & Set of hybrid dirty beams            & sky \& sky \\      
      \hline %
    \end{tabular}
    \tablefoot{This table uses similar conventions as
      Table~\ref{tab:symbols:widefield}. The top part defines symbols
      related to single-dish measurements. The middle part defines symbols
      related to heterogeneous-array measurements. The bottom part defines
      hybrid symbols, which results from combinations of single-dish and
      heterogeneous-array measurements.}
    \label{tab:symbols:shortspacings}
  \end{table}}

\newcommand{\TabSymbolOTF}{%
  \begin{table}
    \centering
    \caption{Definition of the symbols used to explore the influence of
      on-the-fly scanning on the measurement equation.}
    \begin{tabular}{cl}
      \hline %
      \hline %
      \multicolumn{2}{l}{Symbol \& Definition} \\ 
      \hline %
      \dt{}     & Integration time \\
      \asave{}  & Scanned angle averaged during \dt{} \\
      \upave{}  & Spatial frequency averaged during \dt{} \\
      \daslew{} & Angular distance scanned during \dt{} \\
      \vslew{}  & Slew angular velocity of the telescope \\
      \A{}      & Primary beam apodizing function\\
      \Beff{}   & Effective primary beam resulting \\
                & from OTF scanning: $\Beff(\af)=\CONV{\B{}}{\A{}}{\af}$\\
      \wearth{} & Angular velocity of a spatial frequency \\
                & due to Earth rotation \\
      \hline %
    \end{tabular}
    \label{tab:symbols:otf}
  \end{table}}

\newcommand{\FigScales}{%
  \begin{figure}
    \centering{}
    \includegraphics[height=0.75\hsize,angle=270]{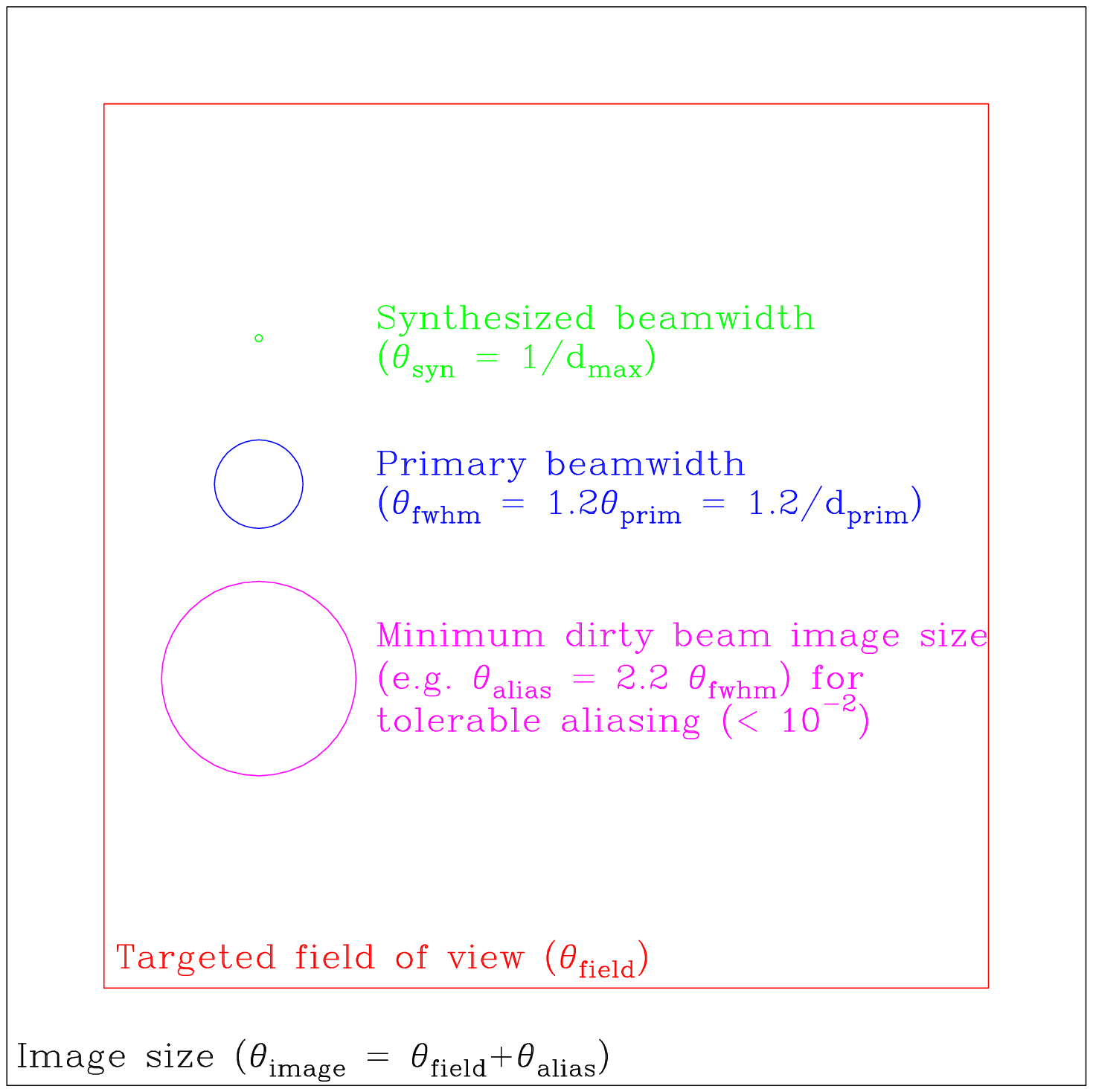}
    \caption{Visualization of the different angular scales relevant to 
      wide-field interferometric imaging. The notion of anti-aliasing scale
      (\Aalias{}) is introduced and discussed in
      Sect.~\ref{sec:resampling}.}
    \label{fig:scales}
  \end{figure}}

\newcommand{\FigPrinciple}{%
  \begin{figure*}
    \centering{}
    \includegraphics[width=\hsize]{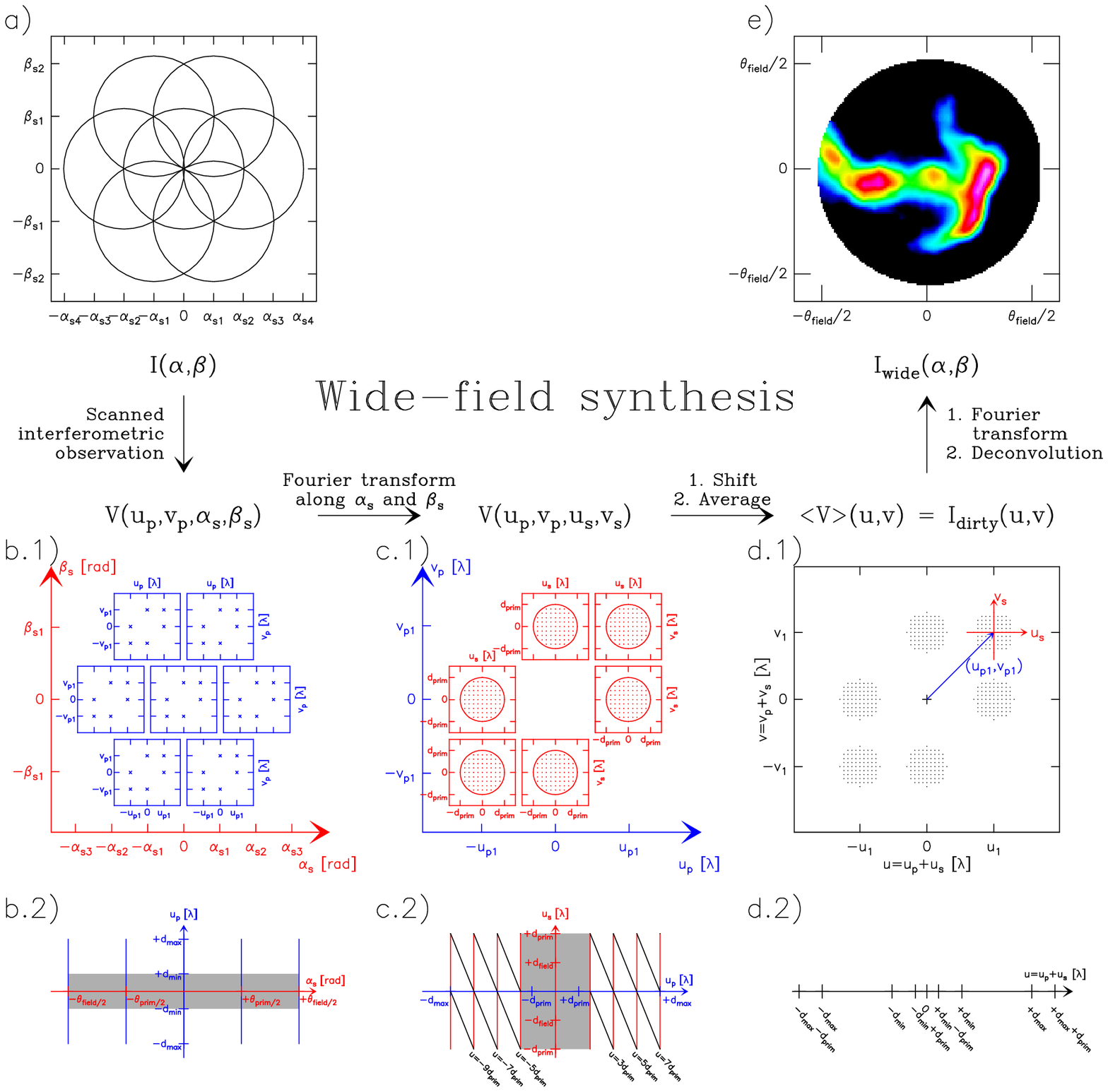}
    \caption{Illustration of the principles of wide-field synthesis, which
      enables us to image wide-field interferometric observations. The top
      row displays the sky plane. The middle row displays the 4-dimensional
      visibility space and the bottom row shows 2-dimensional cuts of this
      space at different stages of the processing. In panels b) to d), the
      scanned dimensions (\as{} and \us{}) are displayed in blue while the
      phased spatial scale dimensions (\up{}) are displayed in red and the
      spatial scale dimensions (\uf{}) of the final wide-field $uv$ plane
      are displayed in black. The grey zones of panels b.2) and c.2) show
      the regions of the visibility space without measurements (missing
      short-spacings). In detail, panel a) shows a possible scanning
      strategy of the sky to measure the unknown brightness distribution at
      high angular resolution: For simplicity it is here just a 7-field
      mosaic. Panel b.1) and b.2) sketch the space of measured
      visibilities: The $uv$ plane at each of the 7 measured sky positions
      is displayed as a blue square box in panel b.1) and a blue vertical
      line in panel b.2). For simplicity, only 6 visibilities are plotted
      in panel b.1). Panels c.1) and c.2) sketch the space of synthesized
      visibilities after Fourier transform of the measured visibilities
      along the scanned coordinate (\as{}): At each measured spatial
      frequency \up{} (displayed on the blue axes) is associated one space
      of synthesized wide-field spatial frequencies displayed as one of the
      red squares in panel c.1) and the red vertical lines in panel c.2).
      The wide-field spatial scales are synthesized 1) on a grid whose cell
      size is related to the total field of view of the observation and 2)
      only inside circles whose radius is the primary diameter of the
      interferometer antennas. Panels d.1) and d.2) display the final,
      wide-field $uv$ plane. This plane is built by application of the
      shift-and-average operator along the black lines on panel c.2), lines
      that display the region of constant \uf{} spatial frequency in the
      (\up,\us) space.  Standard inverse Fourier transform and
      deconvolution methods then produce a wide-field distribution of sky
      brightnesses as shown in panel e).}
    \label{fig:principle}
  \end{figure*}}

\newcommand{\FigPrimaryBeams}{%
  \begin{figure}
    \centering{} \includegraphics[height=\hsize,angle=270]{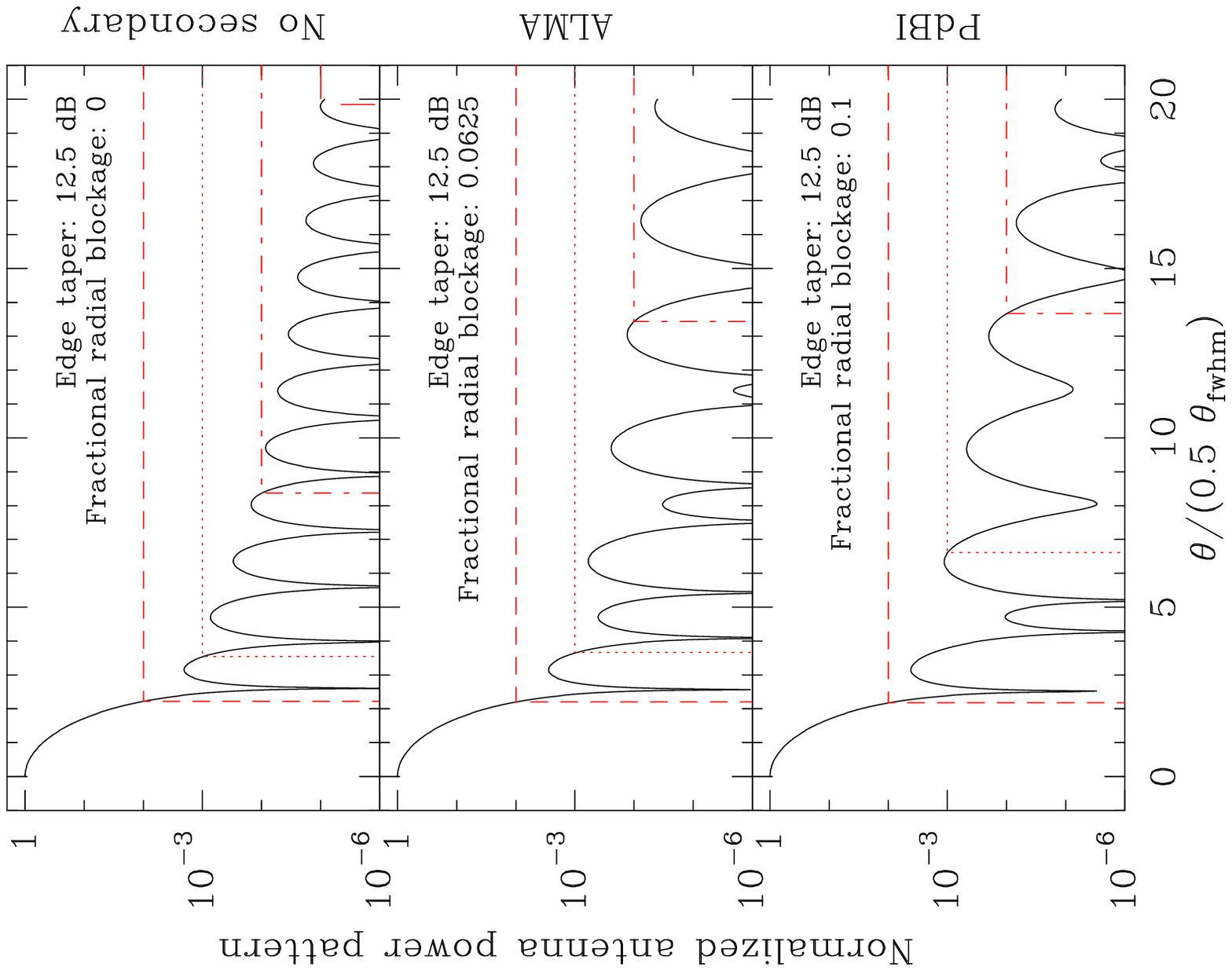}
    \caption{Simple models of the antenna power patterns as a function of
      the sky angle in units of half the primary beam FWHM ($\Afwhm$). In
      the 3 cases shown, the illumination is Gaussian with an edge taper of
      12.5~dB but 3 different ratios of the secondary-to-primary
      diameters (\ie{} \fb{}, the antenna blockage factors) are
      considered~\citep[see \eg{}][chapter 6]{goldsmith98}. The middle and
      bottom panels respectively model ALMA and PdBI antennas. The red
      lines define the minimum angular sizes for which the antenna power
      pattern is less than a given fraction.}
    \label{fig:primbeam}
  \end{figure}}

\newcommand{\FigTextbook}{%
  \begin{figure}
    \centering{}
    \includegraphics[height=\hsize,angle=270]{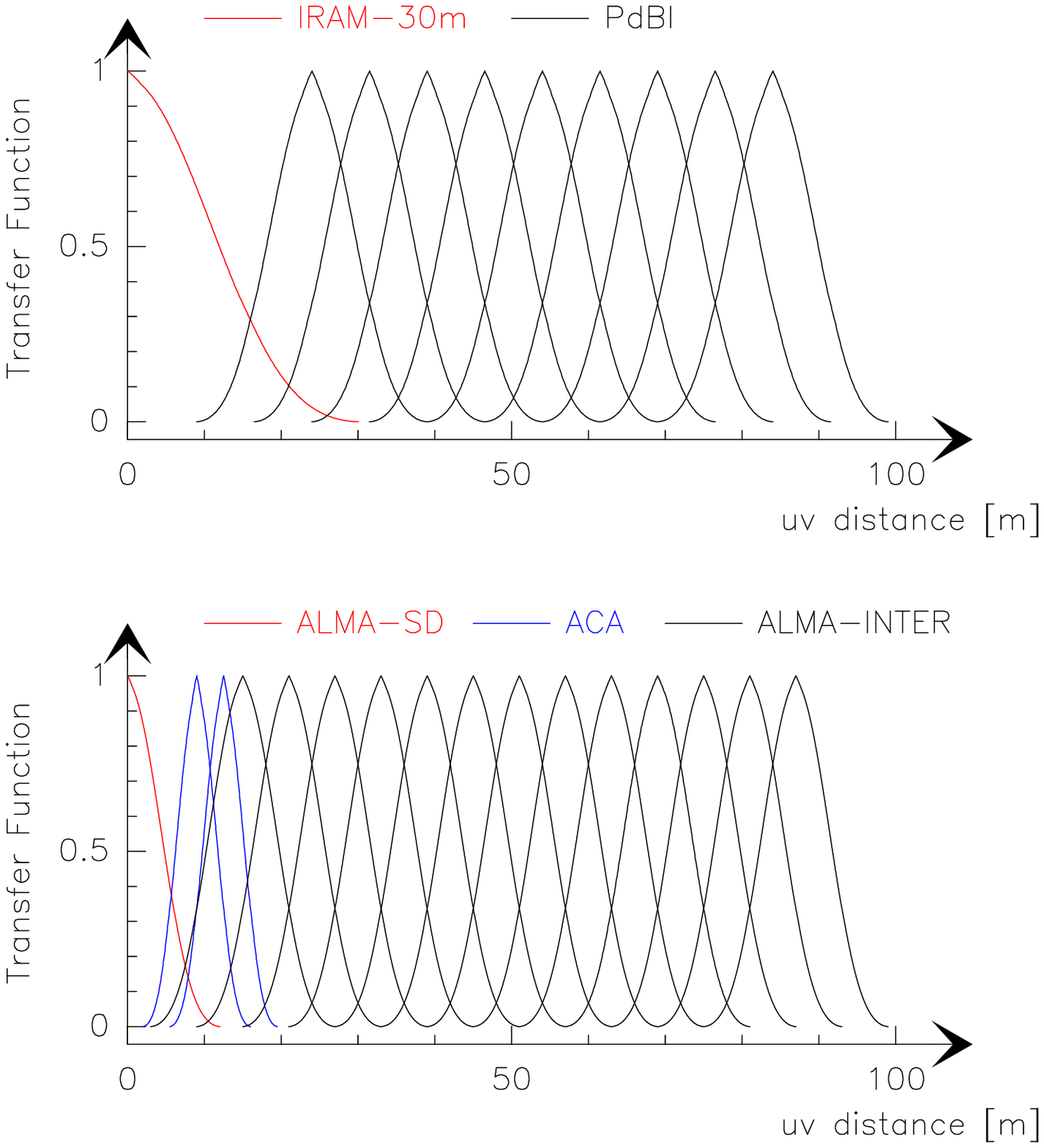}
    \caption{Sketches of the natural weighting of the 
      synthesized wide-field visibilities. Each measured spatial frequency
      will produce wide-field spatial frequencies apodized by the transfer
      function (\Bft{}) centered on the measured spatial frequency. The
      used transfer function depends on the telescopes used, explaining why
      wide-synthesis naturally handles the short spacing either from a
      single-dish antenna or from a heterogeneous array. The synthesized
      visibilities in the overlapping regions will then be averaged.  Two
      textbook examples are illustrated: 1) the combination of data from
      the IRAM-30m single-dish (red transfer function) and from the Plateau
      de Bure Interferometer (black transfer functions) at the top; and 2)
      the combination of data from ALMA 12m-antennas used either in
      single-dish mode (red transfer function), in interferometric mode
      (black transfer functions) and of data from the ACA 7m-antennas (blue
      transfer functions) at the bottom. The minimum $uv$ distances
      measured by each interferometer were set from the minimum possible
      distance between antennas (24~m for PdBI, 15~m for ALMA and 9~m for
      ACA).}
    \label{fig:textbook}
  \end{figure}}

\newcommand{\FigPathLength}{%
  \begin{figure}
    \centering{}
    \includegraphics[height=\hsize,angle=270]{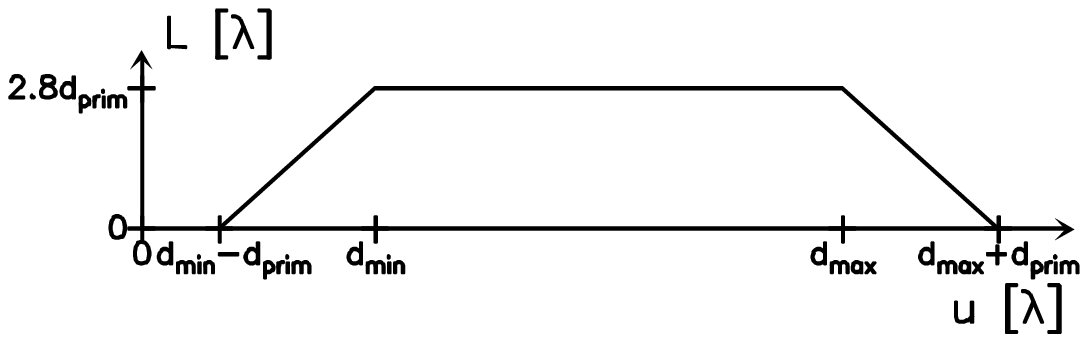}
    \caption{Length of the averaging linepaths displayed as black lines in
      panel c.2) of Fig.~\ref{fig:principle}, as a function of the spatial
      scale in the final, wide-field $uv$ plane. In the case of a
      continuous sampling of \up{} between \dmin{} and \dmax{}, these
      quantities can be interpreted as the number of measures that
      contribute to the estimate of \Ift{}(\uf{}).}
    \label{fig:pathlength}
  \end{figure}}

\newcommand{\FigEffBeam}{%
  \begin{figure}
    \centering
    \includegraphics[height=\hsize,angle=270]{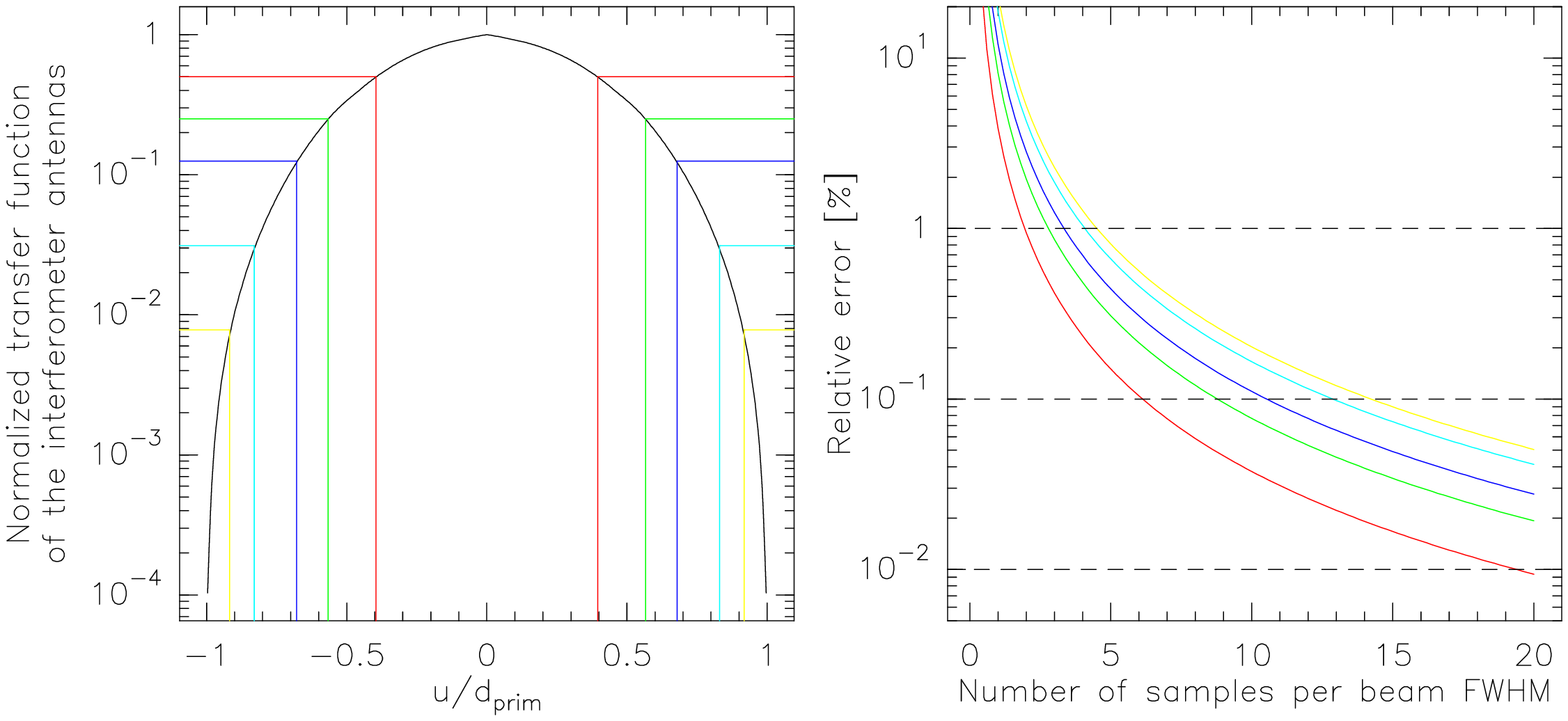}
    \caption{Assessement of the relative error implied by the use of the
      true primary beam instead of the effective primary beam when
      analyzing interferometric on-the-fly data sets. \emph{Left:} Inverse
      Fourier transform of interferometer primary beam, \Bft{} (\ie{} the
      autocorrelation of the antenna illumination).  \emph{Right:} Relative
      error as a function of sampling rate of the primary beam. The curves
      of different colors show the results at different normalized $uv$
      distances ($u/\dprim$) from the center of \Bft{}.}
    \label{fig:EffBeam}
  \end{figure}}

\begin{document}

\title{Revisiting the theory of interferometric wide-field synthesis}
\author{J. Pety\inst{1,2}%
  \and N. Rodr\'iguez-Fern\'andez\inst{1}}

\offprints{\email{pety@iram.fr}}

\institute{IRAM, 300 rue de la Piscine, 38406 Grenoble cedex, France.\\
  \email{pety@iram.fr, rodriguez@iram.fr} \and LERMA, UMR 8112, CNRS and
  Observatoire de Paris, 61 avenue de l'Observatoire, 75014 Paris, France.}

\date{Received , 2009; accepted , 2009}

\abstract %
{After several generations of interferometers in radioastronomy, wide-field
  imaging at high angular resolution is today a major goal for trying to
  match optical wide-field performances.} %
{All the radio-interferometric, wide-field imaging methods currently belong
  to the mosaicking family. Based on a 30 years old, original idea from
  Ekers \& Rots, we aim at proposing an alternate formalism.} %
{Starting from their ideal case, we successively evaluate the impact of the
  standard ingredients of interferometric imaging, \ie{} the sampling
  function, the visibility gridding, the data weighting, and the processing
  of the short spacings either from single-dish antennas or from
  heterogeneous arrays. After a comparison with standard nonlinear
  mosaicking, we assess the compatibility of the proposed processing with
  1) a method of dealing with the effect of celestial projection and 2) the
  elongation of the primary beam along the scanning
  direction when using the on-the-fly observing mode.} %
{The dirty image resulting from the proposed scheme can be expressed as a
  convolution of the sky brightness distribution with a set of wide-field
  dirty beams varying with the sky coordinates.  The wide-field dirty beams
  are locally shift-invariant as they do not depend strongly on position on
  the sky: Their shapes vary on angular scales typically larger or equal to
  the primary beamwidth. A comparison with standard nonlinear mosaicking
  shows that both processing schemes are not mathematically equivalent,
  though they both recover the sky brightness. In particular, the weighting
  scheme is very different in both methods. Moreover, the proposed scheme
  naturally processes the short spacings from both single-dish antennas and
  heterogeneous arrays.  Finally, the sky gridding of the measured
  visibilities, required by the proposed scheme, may potentially save large
  amounts of hard-disk space and cpu processing power over mosaicking when
  handling data sets acquired with
  the on-the-fly observing mode.} %
{We propose to call this promising family of imaging methods
  \emph{wide-field synthesis} because it explicitly synthesizes
  visibilities at a much finer spatial frequency resolution than the one
  set by the diameter of the interferometer antennas.} %

\keywords{techniques: wide-field - techniques: interferometric -
  methods: data analysis - methods: image processing} %

\maketitle{} %

\section{Introduction}

The instantaneous field of view of an interferometer is naturally limited
by the primary beam size of the individual antennas. For the ALMA
12m-antennas, this field of view is $\about9''$ at 690~GHz and $\about
27''$ at 230~GHz.  The astrophysical sources in the (sub)-millimeter domain
are often much larger than this, but still structured on much smaller
angular scales.  Interferometric wide-field techniques enable us to fully
image these sources at high angular resolution.  These techniques first
require an observing mode that in one way or another scans the sky on
spatial scales larger than the primary beam.  The most common observing
mode in use today, known as stop-and-go mosaicking, consists in repeatedly
observing sky positions typically separated by half the primary beam size.
The improvement of the tracking behavior of modern antennas now leads
astronomers to consider on-the-fly observations, with the antennas slewing
continuously across the sky. The improvements in correlator and receiver
technologies are also leading to techniques that could potentially sample
the antenna focal planes with multi-beam receivers instead of the
single-pixel receivers installed on current interferometers.

The ideal measurement equation of interferometric wide-field imaging is
\begin{equation}
  \V{}(\up,\as) = \DFTint{\ap}{\B{}(\ap-\as)\,\I{}(\ap)}{\up},
  \label{eq:measurement:otf}
\end{equation}
where \V{} is the visibility function of 1) \up{} (the spatial frequency
with respect to the fixed phase center) and 2) \as{} (the scanned sky
angle), \I{} is the sky brightness, and \B{} the antenna power pattern or
primary beam of an antenna of the interferometer~\citep[][chapter
2]{thompson86}. For simplicity, 1) we assume that the primary beam is
independent of azimuth and elevation, and 2) we use one-dimensional
notation without loss of generality. We do not deal with
polarimetry~\citep[see \eg][]{hamaker96,sault96a,sault99} because it is
adds another level of complexity over our first goal here, \ie{} wide-field
considerations. Several aspects make Eq.~\ref{eq:measurement:otf} peculiar
with respect to the ideal measurement equation for single-field
observations. First, the visibility is a function not only of the $uv$
spatial frequency (\up{}) but also of the scanned sky coordinate (\as{}).
Second, Eq.~\ref{eq:measurement:otf} is a mix between a Fourier transform
and a convolution equation. It can be regarded, for example, as the Fourier
transform along the \ap{} dimension of the function,
$B{}(\ap-\as)\,\I{}(\ap)$, of the (\ap,\as) variables. But
Eq.~\ref{eq:measurement:otf} can also be written as the convolution:
\begin{equation}
  \V{}(\up,\as) = \sint{\ap}{\Bg{}(\as-\ap)\,\Ig{}(\ap,\up)},
\end{equation}
\begin{equation}
  \mbox{where} \quad \Bg{}(\as-\ap) \Definition \B{}(\ap-\as)
\end{equation}
\begin{equation}
  \mbox{and} \quad \Ig{}(\ap,\up) \Definition \I{}(\ap)\,\dexp{\ap\up}.
\end{equation}
For each \up{} kept constant, \V{}(\up,\as) is the convolution of \Bg{} and
\Ig{}. Indeed, $\Ig{}(\ap,\up = 0) = \I{}(\ap)$, so we derive
\begin{equation}
  \V{}(\up=0,\as) = \conv{\Bg{}}{\I{}}{\ap}{\as}, 
\end{equation}
\ie{}, the convolution equation for single-dish observations.

\citet{ekers79} were the first to show that the measurement equation
(Eq.~\ref{eq:measurement:otf}) enables us to recover spatial frequencies of
the sky brightness at a much finer $uv$ resolution than the $uv$ resolution
set by the diameter of the interferometer antennas.  Interestingly enough,
the goal of \citet{ekers79} was ``just'' to find a way to produce the
missing short spacings of a multiplying interferometer.  However,
\citet{cornwell88} realized that Ekers \& Rots' scheme has a much stronger
impact, because it explains why an interferometer is able to do wide-field
imaging. \citet{cornwell88} also demonstrated that on-the-fly scanning is
not absolutely necessary to interferometric wide-field imaging. Indeed, the
large-scale information can be retrieved in mosaics of single-field
observations, provided that the sampling of the single fields follows the
sky-plane Nyquist sampling theorem.

As a result, all the information about the sky brightness is coded in the
visibility function. From a data-processing viewpoint, all the current
radio-interferometric wide-field imaging methods~\citep[see,
\eg{},][]{gueth95,sault96b,cornwell93,bhatnagar04,bhatnagar08,cotton08}
belong to the mosaicking family\footnote{In the rest of this paper,
  stop-and-go mosaicking refer to the observing mode, while mosaicking
  alone refer to the imaging family.} pioneered by \citet{cornwell88}.  In
this family, the processing starts with Fourier transforming
\V{}(\up{},\as{}) along the \up{} dimension (\ie{} at constant \as{}) to
produce a set of single-field dirty images before linearly combining them
and forming a wide-field dirty image. In this paper, we propose an
alternate processing, which starts with a Fourier transform of
\V{}(\up{},\as{}) along the \as{} dimension (\ie{} at constant \up{}). We
show how this explicitely synthesizes the spatial frequencies needed to do
wide-field imaging, which are linearly combined to form a ``wide-field $uv$
plane'', \ie{}, one $uv$-plane containing all the spatial frequency
information measured during the wide-field observation.  Inverse Fourier
transform will produce a dirty image, which can then be deconvolved using
standard methods. The existence of two different ways to extract the
wide-field information from the visibility function raises several
questions: Are they equivalent? What are their relative merits?

We thus aim at revisiting the mathematical foundations of wide-field
imaging and deconvolution. Sections~\ref{sec:notations}
to~\ref{sec:mosaicking} propose the new algorithm, which we call wide-field
synthesis: Section~\ref{sec:notations} first defines the notations and it
then lays out the basic concepts used throughout the paper.
Section~\ref{sec:ekers-rots} states the steps needed to go beyond the Ekers
\& Rots scheme and explores the consequences of incomplete sampling of both
the $uv$ and sky planes.  Section~\ref{sec:gridding} discusses the effects
of gridding by convolution and regular resampling.
Section~\ref{sec:deconvolution} describes how to influence the dirty beam
shapes and thus the deconvolution.  Section~\ref{sec:short-spacings} states
how to introduce short spacings measured either from a single-dish antenna
or from heterogeneous interferometers. Section~\ref{sec:mosaicking}
compares the proposed wide-field synthesis algorithm with standard
nonlinear mosaicking. Some detailed demonstrations are factored out in
Appendix~\ref{sec:demo} to enable an easier reading of the main paper,
while ensuring that interested readers can follow the demonstrations.
Appendices~\ref{sec:projection} and~\ref{sec:otf} then explain how the
wide-field synthesis algorithm can cope with non-ideal effects:
Appendix~\ref{sec:projection} discusses how at least one standard way to
cope with sky projection problems is compatible with the wide-field
synthesis algorithm.  Appendix~\ref{sec:otf} explores the consequences of
using the on-the-fly observing mode. Finally, we assume good familiarity
with single-field imaging in various places. We refer the reader to
well-known references: \eg{} chapter 6 of \citet{thompson86}
and~\citet{sramek89}.

\vspace{-\medskipamount}

\section{Notations and basic concepts}
\label{sec:notations}

\subsection{Notations}

In this paper, we use the \citet{bracewell00}'s notation to display the
relationship between a function \I{}(\af) and its direct Fourier transform
\Ift{}(\uf), \ie{},
\begin{equation}
  \FTop{\I{}(\af)}{\Ift{}(\uf)}{}{},
\end{equation}
where (\af,\uf) is the couple of Fourier conjugate variables. We also use
the following sign conventions for the direct and inverse Fourier
transforms
\begin{equation}
  \Ift{}(\uf) \Definition \DFTint{\af}{\I{}(\af)}{\uf} 
\end{equation}
and
\begin{equation}
  \I{}(\af) \Definition \IFTint{\uf}{\Ift{}(\uf)}{\af}.
\end{equation}
As \V{} is a function of two independent quantities (\up{} and \as{}), the
Fourier transform may be applied independently on each dimension, while the
other dimension stays constant. Several additional conventions are used to
express this. First, we introduce a specific notation to state that either
the first or the second dimension stays constant:
\begin{equation}
  \V{\up}(\as) \Definition \V{}(\up=\emr{constant},\as),
\end{equation}
and
\begin{equation}
  \V[\as]{}(\up) \Definition \V{}(\up,\as=\emr{constant}).
\end{equation}
Second, we use a bottom/top line to derive the notation of the Fourier
transform along the first/second dimension from the notation of the
original function. Third, on the Fourier transform sign (\ie{} $\supset$),
we explicitly state the dimension along which the Fourier transform is
computed. For instance, if \Dw{} is a function of (\ap{},\as{}), then the
Fourier transform of \Dw{} along the first dimension is expressed as
\begin{equation}
  \FTop{\Dw{}(\ap,\as)}{\Dwftnn{}(\up,\as)}{\ap}{\up},
\end{equation}
while the Fourier transform of \Dw{} along the second dimension is
expressed as
\begin{equation}
  \FTop{\Dw{}(\ap,\as)}{\Dwnnft{}(\ap,\us)}{\as}{\us}.
\end{equation}
We also use a more compact notation when doing the Fourier transform on
both dimensions simultaneously, \ie{},
\begin{equation}
  \FTop{\Dw{}(\ap,\as)}{\Dwftft{}(\up,\us)}{(\ap,\as)}{(\up,\us)}.
\end{equation}
Finally, the convolution of two functions $G$ and \V{} is noted and defined
as
\begin{equation}
  \CONV{G}{\V{}}{u} \Definition \conv{G}{\V{}}{v}{u}.
  \label{eq:def:conv}
\end{equation}

\TabSymbolWideField{} %

For reference, Table~\ref{tab:symbols:widefield} summarizes the definitions
of the symbols used most throughout the paper. With the one-dimensional
notation used throughout the paper, the number of planes quoted directly
gives the number of associated dimensions of the symbols. Generalization to
images would require a doubling of the number of planes/dimensions.
Table~\ref{tab:scales} defines the $uv$ and angular scales that are
relevant to wide-field interferometric imaging, and Fig.~\ref{fig:scales}
sketches the different angular scales. Each angular scale ($\theta$) is
related to a $uv$ scale ($d$) through $\theta = 1/d$, where $\theta{}$ and
$d$ are measured in radians and in units of \wavelength{} (the wavelength
of the observation).  In the rest of the paper, we explicitely distinguish
between $\Aprim \Definition 1/\dprim$, the angular scale associated to the
diameter of the interferometer antennas, and \Afwhm{}, the full width at
half maximum of the primary beam. The relation between \Aprim{} and
\Afwhm{} depends on the illumination of the receiver feed by the antenna
optics. In radio astronomy, we typically have $\Afwhm \about
1.2\,\Aprim$~\citep[see \eg{}][chapter 6]{goldsmith98}. Finally, the notion
of anti-aliasing scale (\Aalias{}) is introduced and discussed in
Sect.~\ref{sec:resampling}.

\TabScales{} %
\FigScales{} %

\subsection{Basic concepts}

\FigPrinciple{} %

Figure~\ref{fig:principle} illustrates the principles underlying 1) the
setup to get interferometric wide-field observations and 2) our proposition
to process them. For simplicity, we display the minimum possible complexity
without loss of generality. The top row displays the sky plane. The middle
row represents the 4-dimensional measurement space at different stages of
the processing. As it is difficult to display a 4-dimensional space on a
sheet of paper, the bottom row shows 2-dimensional cuts of the measurement
space at the same processing stages.

\vspace{-\medskipamount}

\subsubsection{Observation setup and measurement space}

Panel a) displays the sky region for which we aim for estimating the sky
brigthness, \I{}(\af{}). The field of view of an interferometer observing
in a given direction of the sky has a typical size set by the primary beam
shape. In our example, this is illustrated by any of the circles whose
diameter is $\Aprim$. As we aim at observing a wider field of view, \eg{}
\Afield{}, the interferometer needs to scan the targeted sky field.  We
assume that we scan through stop-and-go mosaicking, ending up with a
7-field mosaic.

After calibration, the output of the interferometer is a visibility
function, \V{}(\up{},\as{}), whose relation to the sky brightness is given
by the measurement equation (Eq.~\ref{eq:measurement:otf}). Panel b.1)
shows the measurement space as a mosaic of single-field $uv$ planes: The
$uv$ plane coverage of each single-field observation is displayed as a blue
sub-panel at the sky position where it has been measured and which is
featured by the red axes. We assume 1) that the interferometer has only 3
antennas and 2) that only a single integration is observed per sky
position. This implies only 6 visibilities per single-field $uv$ plane.  In
panel b.2), the $uv$ planes at constant \as{} are displayed as the blue
vertical lines. The measured spatial frequencies belong to the
$[-\dmax,-\dmin]$ and $[+\dmin,+\dmax]$ ranges, where \dmin{} and \dmax{}
are respectively the shortest and longest measured baseline length.
\dmin{} is related to the minimum tolerable distance between two antennas
to avoid collision. Here, we chose $\dmin \about 1.5\,\dprim$. The grey
zone between $-\dmin$, and $+\dmin$ displays the missing short spacings.

\vspace{-\smallskipamount}

\subsubsection{Processing by explicit synthesis of the wide-field spatial frequencies}

All the information about the sky brightness, \I{}(\af{}), is somehow coded
in the visibility function, \V{}(\up{},\as{}). The high spatial frequencies
(from \dmin{} to \dmax{}) are clearly coded along the \up{} dimension.  The
uncertainty relation between Fourier conjugate quantities also implies that
the typical spatial frequency resolution along the \up{} dimension is only
\dprim{} because the field of view of a single pointing has a typical size
of \Aprim{}. However, wide-field imaging implies measuring all the spatial
frequencies with a finer resolution, $\dfield = 1/\Afield$. The missing
information must then be hidden in the \as{} dimension.

In Sect.~\ref{sec:ekers-rots}, we show that Fourier transforming the
measured visibilities along the \as{} dimension (\ie{} at constant \up{})
can synthesize the missing spatial frequencies, because the \as{} dimension
is sampled from $-\Afield/2$ to $+\Afield/2$, implying a typical
spatial-frequency resolution of the \us{} dimension equal to \dfield{}.
Conversely, the \as{} dimension is probed by the primary beams with a
typical angular resolution of \Aprim{}, implying that the \us{} spatial
frequencies will only be synthesized inside the $[-\dprim,+\dprim]$ range.
Panels c.1) and c.2) illustrate the effects of the Fourier transform of
\V{}(\up,\us) along the \as{} dimension, in 4 and 2 dimensions,
respectively. The red subpanels or vertical lines display the \us{} spatial
frequencies around each constant \up{} spatial frequency.

In panels d.1) and d.2) (\ie{} after the Fourier transform along the \as{}
dimension), \Vft{}(\up,\us{}) contains all the measured information about
the sky brightness in a spatial frequency space. However, the information
is ordered in a strange and redundant way. Indeed, we show that
\Vft{}(\up,\us) is linearly related to \Ift{}(\up+\us). To first order, the
information about a given spatial frequency \uf{} is stored in all the
values of \Vft{}(\up,\us) which verifies $\uf=\up+\us$ (black lines on
panel c.2).
  
A shift operation will reorder the spatial scale information and averaging
will compress the redundancy (illustrated by the halving of the number of
the space dimensions). The use of a shift-and-average operator thus
produces a final $uv$ plane containing all the spatial scale information to
image a wide field in an intuitive form. We thus call this space the
\emph{wide-field $uv$ plane}. Panels d.1) and d.2) display this space,
where the minimum relevant spatial frequency is related to the total field
of view, while the maximum one is related to the interferometer resolution.

Sections~\ref{sec:ekers-rots} and~\ref{sec:gridding} show that applying the
shift-and-average operator to \Vft{} produces the Fourier transform of a
dirty image, which is a local convolution of the sky brightness by a slowly
varying dirty beam. As a result, inverse Fourier transform of
\average{\Vft{}} and deconvolution methods will produce a wide-field
distribution of sky brightness as shown in panel e) at the top right of
Fig.~\ref{fig:principle}.

\vspace{-\medskipamount}

\section{Beyond the Ekers \& Rots scheme}
\label{sec:ekers-rots}

In the real world, the visibility function is not only sampled, but this
sampling is incomplete for two main reasons. 1) The instrument has a finite
spatial resolution, and the scanning of the sky is limited, implying that
the sampling in both planes has a finite support. 2) The $uv$ coverage and
the sky-scanning coverage can have holes caused either by intrinsic
limitations (\eg{} lack of short spacings or small number of baselines) or
by acquisition problems (implying data flagging). The incomplete sampling
makes the mathematics on the general case complex. We thus start with the
ideal case where we assume that the visibility function is continuously
sampled along the \up{} and \as{} dimension. We then look at the general
case.

\subsection{Ideal case: Infinite, continuous sampling}
\label{sec:ideal}

Starting from the measurement equation~\ref{eq:measurement:otf},
\citet{ekers79} first demonstrated (see Sect.~\ref{sec:demo:er})
that\footnote{The convolution theorem, which states that the Fourier
  transform of the convolution of two functions is the product of the
  Fourier transform of both individual functions, is a special case for
  Eq.~\ref{eq:ekers-rots:1}: It can be recovered by setting $\up=0$.
  Indeed, as already mentioned in the introduction, the ideal measurement
  equation~\ref{eq:measurement:otf} can be interpreted as a convolution
  with an additional phase term. By Fourier transforming along the \as{}
  dimension, the convolution translates into a product of Fourier
  transforms \Bft{} and \Ift{}, while the phase term translates into a
  shift of coordinates: $\up+\us$.}
\begin{equation}
  \forall (\up,\us), \quad \Vft{\up}(\us) = \Bft{}(-\us)\,\Ift{}(\up+\us).
  \label{eq:ekers-rots:1}
\end{equation}
For each constant \up{} spatial frequency, the Fourier transform thus
synthesizes a function, \Vft{\up}(\us), which is simply related to
\Ift{}(\up+\us), the Fourier components of the sky brightness around \up{}.
\Vft{}(\up,\us) is only defined in the $[-\dprim,+\dprim]$ interval along
the \us{} dimension because \Bft{}(-\us{}) is itself only defined inside
this interval, since \Bft{}(-\us{}) is the autocorrelation of the antenna
illumination.

We search to derive a single estimate of the Fourier components
$\Ift{}(\uf)$ of the sky brightness. Equation~\ref{eq:ekers-rots:1}
indicates that the fraction $\Vft{}(\up,\us)/\Bft{}(-\us)$ gives us an
estimate of \Ift{}(\uf) for each couple $(\up,\us)$ that satisfies
$\uf=\up+\us$.  However, the information about \Ift{} is strangely ordered.
There are two possible ways to look at this ordering. 1) Starting from the
measurement space, the Ekers \& Rots scheme synthesizes frequencies around
each \up{} measure inside the interval $[\up-\dprim, \up+\dprim]$ at the
\dfield{} spatial frequency resolution. 2) Starting from our goal, we want
to estimate \Ift{} at a given spatial frequency \uf{} with a \dfield{}
spatial frequency resolution. We thus search for all the couples
$(\up,\us)$ satisfying $\uf=\up+\us$, which are displayed in panel c.2) of
Fig.~\ref{fig:principle} as the diagonal black lines. It immediately
results that 1) there are several estimates of \Ift{} for each spatial
frequency \uf{} and 2) the number of estimates varies with \uf{}. We can
average them to get a better estimate of \Ift{}(\uf).

This last viewpoint thus suggests averaging in the (\up,\us) space along
linepaths defined by $\uf=\up+\us$. Such an operator can mathematically be
defined as
\begin{equation}
  \average{F}(\uf) \!\Definition \!\!\!\dint{\up}{\us}{\!\!\!\!\!\!\linepath\,\Wu{}(\up,\us)\,F(\up,\us)},
  \label{eq:average:1}
\end{equation}
where $F$ is the function to be averaged and \Wu{} is a normalized
weighting function. Using the properties of the Dirac function, we can
reduce the double integral to
\begin{equation}
  \average{F}(\uf) = \sint{\up}{\Wu{}(\up,\uf-\up)\,F(\up,\uf-\up)}.
  \label{eq:average:2}
\end{equation}
In this equation, we easily recognize a \emph{shift-and-average operator}.
The normalized weighting function plays a critical role in the following
formalism, and we propose clever ways to define \Wu{} in
Sect.~\ref{sec:deconvolution}. In the ideal case studied here, \Wu{} can be
defined as
\begin{description}
\item[$\Wu{}(\up,\us) \Definition 1/(2\sqrt{2}\,\dprim)$] for $\us
  [-\dprim,+\dprim]$,
\item[$\Wu{}(\up,\us) \Definition 0$] for other values of \us{}.
\end{description}
In other words, we have just normalized the integral by the constant length
($2\sqrt{2}\,\dprim$) of the averaging linepath.

\subsubsection{Wide-field dirty image, dirty beam and image-plane 
  measurement equation}

Section~\ref{sec:real} shows that the incomplete sky and $uv$ sampling
forbid us to apply the shift-and-average operator to the
$\Vft{}(\up,\us)/\Bft{}(-\us)$ function. To guide us in this general case,
we thus explore the consequences of applying this operator to \Vft{} in the
ideal case.  It is easy to demonstrate that the result is the Fourier
transform of a dirty image, \ie{},
\begin{equation}
  \Iwfft{}(\uf) = \average{\Vft{}}(\uf).
\end{equation}
Indeed, substituting $\average{\Vft{}}(\uf)$ with the help of
Eqs.~\ref{eq:average:2} and~\ref{eq:ekers-rots:1} and taking the inverse
Fourier transform, we get
\begin{equation}
  \Iwfft{}(\uf) = \Dwft{}(\uf) \,\Ift{}(\uf),
\end{equation}
\begin{equation}
  \mbox{with} \quad \Dwft{}(\uf) \Definition \sint{\up}{\Wu{}(\up,\uf-\up)\,\Bft{}(\up-\uf)}.
  \label{eq:dirty:uv}
\end{equation}
Here, \Iwf{} conforms to the usual idea of dirty image, \ie{}, the
convolution of a dirty beam by the sky brightness:
\begin{equation}
  \Iwf{}(\af) = \CONV{\Dw{}}{\I{}}{\af}.
\end{equation}
In contrast to the usual situation for single-field observations, the mix
between a Fourier transform and a convolution of
Eq.~\ref{eq:measurement:otf}, associated with the specific
processing\footnote{\ie{} direct Fourier transform along the \as{}
  dimension, shift-and-average to define a final wide-field $uv$ plane, and
  inverse Fourier transform.} changes the image-plane measurement equation
from a convolution of a dirty beam with the product $\B{}\,\I{}$ to a
convolution of a dirty beam with \I{}. \emph{The dependency on the primary
  beam is still there. It is just transferred from a product of the sky
  brightness distribution into the definition of the dirty beam.}

\subsubsection{Summary and interpretation}

In summary, a theoretical implementation of wide-field synthesis implies
\begin{enumerate}
\item the possibility of Fourier transforming the visibility function along
  the \as{} dimension (\ie{} at constant \up{}), which gives us a set of
  synthesized $uv$ planes;
\item the possibility of shifting-and-averaging these synthesized $uv$
  planes to build the final, wide-field $uv$ plane containing all the
  available information.
\end{enumerate}
Using those tools, we are able to write the wide-field image-plane
measurement equation as a convolution of a wide-field dirty beam (\Dw{}) by
the sky brightness (\I{}), \ie{},
\begin{equation}
  \Iwf(\af) = \sint{\aff}{\Dw{}(\af-\aff)\,\I{}(\aff)}.
  \label{eq:ideal:meas:im}
\end{equation}
We can write a convolution equation in this ideal case because the
wide-field response of the instrument is shift-invariant; \ie{}, \Dw{} only
depends on differences of the sky coordinates.

It is well-known that for a single-field observation, the dirty beam is the
inverse Fourier transform of the sampling function. The shape of this
sampling function is due to the combination of aperture synthesis (the
interferometer antennas give a limited number of independent baselines) and
Earth-rotation synthesis (the rotation of the Earth changes the projection
of the physical baselines onto the plane perpendicular to the instantaneous
line of sight). By analyzing via a Fourier transform, the evolution of the
visibility function with the sky position, the Ekers \& Rots scheme
synthesizes visibilities at spatial frequencies needed to image a larger
field of view than the interferometer primary beam. We thus propose to call
this specific processing: \emph{wide-field synthesis}.

\subsection{General case: Incomplete sampling}
\label{sec:real}

Reality imposes limitations on the synthesis of spatial frequencies.
Indeed, we have already stated that the visibility function is incompletely
sampled both in the $uv$ and sky planes.  To take the sampling effects into
account, we introduce the sampling function \S{}(\up,\as), which is a sum
of Dirac functions at measured positions\footnote{Loosely speaking, the
  sampling function can be thought as a function whose value is 1 where
  there is a measure and 0 elsewhere.}. The sampling function cannot be
factored into the product of two functions, each only acting on one plane.
Indeed, the Earth rotation happening during the source scanning implies a
coupling of both dimensions of the sampling function. In other words, the
$uv$ coverage will vary with the scanned sky coordinate. This leads us to a
shift-dependent situation, precluding us from writing the wide-field
image-plane measurement equation as a true convolution. We nevertheless
search for a wide-field image-plane measurement equation as close as
possible to a convolution because all the inversion methods devised in the
past three decades in radioastronomy are tuned to deconvolve images. The
simplest mathematical way to generalize Eq.~\ref{eq:ideal:meas:im} to a
shift-dependent situation is to write it as
\begin{equation}
  \Iwf(\af) = \sint{\aff}{\Dw{}(\af-\aff,\af)\,\I{}(\aff)}.
  \label{eq:samp:meas:im}
\end{equation}
In this section, we show how the \emph{linear character of the imaging
  process} allows us to do this. Section~\ref{sec:real:ekers-rots} derives
the impact of incomplete sampling on the Ekers \& Rots equation, and
Sect.~\ref{sec:real:uv-plane} derives the wide-field measurement equation
in the $uv$ plane.  Section~\ref{sec:real:interpretation} interprets these
results.

\subsubsection{Effect on the Ekers \& Rots equation}
\label{sec:real:ekers-rots}

The sampled visibility function, \SV{}, is defined as the product of \S{}
and \V{} and \SVft{} its Fourier transform along \as{}, \ie{},
\begin{equation}
  \SV{}(\up,\as) \Definition \S{}(\up,\as)\,\V{}(\up,\as),
\end{equation}
\begin{equation}
  \mbox{and} \quad \FTop{\SV{}(\up,\as)}{\SVft{}(\up,\us)}{\as}{\us}.
\end{equation}
Because \SV{\up} is the product of two functions of \as, we can use the
convolution theorem to show that \SVft{\up} is the convolution of \Sft{\up}
by \Vft{\up}, \ie{},
\begin{equation}
  \SVft{\up}(\us) = \conv{\Sft{\up}}{\Vft{\up}}{\uss}{\us}.
  \label{eq:samp:conv}
\end{equation}
By replacing $\Vft{\up}$ with the help of the Ekers \& Rots relation
(Eq.~\ref{eq:ekers-rots:1}), we derive
\begin{equation}
  \SVft{\up}(\us) = \sint{\uss}{\Sft{\up}(\us-\uss)\,\Bft{}(-\uss)\,\Ift{}(\up+\uss)}.
  \label{eq:ekers-rots:2}
\end{equation}
As \Bft{} is bounded inside the $[-\dprim,+\dprim]$ interval,
\SVft{\up}(\us{}) is a \emph{local} average, weighted by
$\Sft{\up}(\us-\uss)\,\Bft{}(-\uss)$, of \Ift{}(\up+\uss) around the \up{}
spatial frequency.

As expected, we recover Eq.~\ref{eq:ekers-rots:1} for the ideal case
(\ie{}, infinite, continuous visibility function) because then
$\Sft{\up}(\us-\uss) = \dirac{}(\us-\uss)$. A more interesting case arises
when the visibility function is continuously sampled over a limited sky
field of view, \ie{},
\begin{eqnarray}
  \forall \up, &\quad& \S{}(\up,\as) = 1 \quad \mbox{if} \quad \abs{\as} \le \Afield/2,\\
  \forall \up, &\quad& \S{}(\up,\as) = 0 \quad \mbox{if} \quad \abs{\as} > \Afield/2.
\end{eqnarray}
After Fourier transform this gives
\begin{equation}
  \forall \up, \quad \Sft{}(\up,\us) = \frac{1}{\dfield}\,\sinc\!\paren{\frac{\us}{\dfield}}.
\end{equation}
In this case, the local average of the sky brightness Fourier components
happens on a typical $uv$ scale equal to \dfield. However, the \sinc{}
function is known to decay only slowly. Some observing strategy (\eg{}
quickly observing outside the edges of the targeted field of view to
provide a bandguard) could be considered to apodize the sky-plane
dependence of the sampling function, resulting in faster decaying \Sft{}
functions, hence in less mixing of the wide-field spatial frequencies.

\subsubsection{$uv$-plane wide-field measurement equation}
\label{sec:real:uv-plane}

Because we aim at estimating the Fourier component of \Ift{}, we introduce
the following change of variables $\uff \Definition \up+\uss$ and $d\uff =
d\uss$, to derive
\begin{equation}
  \SVft{\up}(\us) = \!\!\sint{\uff}{\!\!\!\!\Sft{\up}(\up+\us-\uff)\,\Bft{}(\up-\uff)\,\Ift{}(\uff)}.
  \label{eq:ekers-rots:3}
\end{equation}
We then shift-and-average \SVft{}(\up,\us) to build the Fourier transform
of a wide-field dirty image
\begin{equation}
  \Iwfft{}(\uf) \Definition \average{\SVft{}}(\uf), 
  \quad \mbox{with} \quad 
  $\uf = \up+\us$.
\end{equation}
Substituting the shift-and-average operator by its definition and using
Eq.~\ref{eq:ekers-rots:3} to replace \SVft{\up}(\us), we derive\\
\Iwfft{}(\uf)
\begin{equation}
  = \!\!\!\dint{\up}{\uff}{\!\!\!\!\!\!\Wu{}(\up,\uf-\up)\,\Sft{}(\up,\uf-\uff)\,\Bft{}(\up-\uff)\,\Ift{}(\uff)}.
  \label{eq:samp:dirty:1}
\end{equation}
This $uv$-plane wide-field measurement equation can be written as
\begin{equation}
  \Iwfft{}(\uf) = \sint{\uff}{\Dwftft{}(\uff,\uf-\uff)\,\Ift{}(\uff)},
  \label{eq:samp:meas:uv}
\end{equation}
if we enforce the following equality
\begin{equation}
  \Dwftft{}(\uff,\uf-\uff) \Definition \!\!\!\sint{\up}{\!\!\!\!\!\Wu{}(\up,\uf-\up)\,\Sft{}(\up,\uf-\uff)\,\Bft{}(\up-\uff)}.
  \label{eq:dirty-beam}
\end{equation}
This is one way to define \Dwftft{}, which is convenient though unusual.
It is implicit in this definition that we need to make a change of variable
($\ufff=\uf-\uff$) to derive
\begin{equation}
  \Dwftft{}(\uff,\ufff) \Definition \!\!\!\sint{\up}{\!\!\!\!\!\Wu{}(\up,\uff+\ufff-\up)\,\Sft{}(\up,\ufff)\,\Bft{}(\up-\uff)}.
\end{equation}
In the following, we use either one or the other definition of \Dwftft{},
depending on convenience.

\subsubsection{Interpretation}
\label{sec:real:interpretation}

Appendix~\ref{sec:demo:samp} demonstrates that the image and $uv$-plane
wide-field measurement equations (Eqs.~\ref{eq:samp:meas:im}
and~\ref{eq:samp:meas:uv}) are equivalent if
\begin{equation}
  \FTop{\Dw{}(\ap,\as)}{\Dwftft{}(\up,\us)}{(\ap,\as)}{(\up,\us)}.
\end{equation}
The image-plane wide-field measurement equation (Eq.~\ref{eq:samp:meas:im})
can be written as
\begin{equation}
  \Iwf(\af) = \CONV{\Dw[\af]{}}{\I{}}{\af}.
\end{equation}
Its interpretation is straightforward: The sky brightness distribution is
convolved with a dirty beam, \Dw{}(\aff,\afff), which varies with the sky
coordinate \afff{}. This raises the question of the rate of change of the
dirty beam with the sky coordinate. This question is addressed in
Sects.~\ref{sec:resampling} and~\ref{sec:deconvolution}.

\section{Gridding by convolution and regular resampling}
\label{sec:gridding}

We want to Fourier transform the raw visibilities along the sky dimension
(\as{}) at some constant value in the \up{} dimension. The raw data,
however, is sampled on an irregular grid in both the $uv$ and sky planes.
We need to grid the measured visibilities in both the $uv$ and the sky
planes before Fourier transformation for different reasons. First, the
gridding in the $uv$ plane will handle the variation in the spatial
frequency as the sky is scanned, \ie{}, the difficulty and perhaps the
impossibility of Fourier-transforming at a completely constant \up{} value.
Second, the gridding along the sky dimension allows the use of Fast Fourier
Transforms. As usual, we grid through convolution and regular resampling.

\subsection{Convolution}

\subsubsection{Definitions}

We first define a gridding kernel that depends on both dimensions,
$\G{}(\uf,\as)$. This gridding kernel can be chosen as the product of two
functions, simplifying the following demonstrations:
\begin{equation}
  \G{}(\up,\as) \Definition \gu{}(\up)\,\ga{}(\as).
  \label{eq:grid:kernel}
\end{equation}
We then define the sampled visibility function gridded in both the $uv$ and
sky planes as
\begin{eqnarray}
  && \SV[\G{}]{}(\up,\as) \Definition \CONV{\G{}}{\SV{}}{\up,\as}\\
  &=& \dint{\upp}{\ass}{\gu{}(\up-\upp)\,\ga{}(\as-\ass)\,\SV{}(\upp,\ass)}.
  \label{eq:grid:conv}
\end{eqnarray}
Finally, when assessing the impact of the gridding on the measurement
equation~\ref{eq:samp:meas:uv}, a new function,
\begin{equation}
  \SB{}(\up,\as,\asss) \Definition \S{}(\up,\as)\,\B{}(\asss-\as),
  \label{eq:samp:def:1}
\end{equation}
and its Fourier transforms naturally appear in the equations. Defining the
following Fourier transform relationships
\begin{equation}
  \FTop{\SB{}(\up,\as,\asss)}{\SBnnft{}(\up,\us,\asss)}{\asss}{\usss},
\end{equation}
and
\begin{equation}
  \FTop{\SBnnft{}(\up,\us,\asss)}{\SBftft{}(\up,\us,\usss)}{\as}{\us},
\end{equation}
we easily derive
\begin{equation}
  \SBnnft{}(\up,\as,\usss) = \S{}(\up,\as)\,\Bft{}(\usss)\,\dexp{\usss\as},
  \label{eq:samp:def:2}
\end{equation}
and
\begin{equation}
  \SBftft{}(\up,\us,\usss) = \Sft{}(\up,\us+\usss)\,\Bft{}(\usss).
  \label{eq:samp:def:3}
\end{equation}
Using these notations, we have before gridding,
\begin{equation}
  \SV{}(\up,\as) = \!\!\DFTint{\ap}{\!\!\!\!\!\SB{}(\up,\as,\ap)\,\I{}(\ap)}{\up},
\end{equation}
and
\begin{equation}
  \Dwftft{}(\uff,\uf-\uff) \!= \!\!\!\sint{\up}{\!\!\!\!\!\Wu{}(\up,\uf-\up)\,\SBftft{}(\up,\uf-\up,\up-\uff)}.
\end{equation}

\subsubsection{Conservation of the wide-field measurement equation}

Appendix~\ref{sec:demo:grid} demonstrates that the wide-field dirty image
is here again the convolution of the sky brightness \I{} by a wide-field
dirty beam \Dw[\af]{} or, in the Fourier plane,
\begin{equation}
  \Iwfft[\G{}](\uf) \Definition \average{\SVft[\G{}]{}}(\uf)
  = \sint{\uff}{\Dwftft[\G{}]{}(\uff,\uf-\uff)\,\Ift{}(\uff)}
\end{equation}
with
\begin{equation}
  \Dwftft[\G{}]{}(\uff,\uf-\uff) \!\Definition
  \!\!\sint{\up}{\!\!\!\!\!\Wu{}(\up,\uf-\up)\,\SBftft[\G{}]{}(\up,\uf-\up,\uff)},
  \label{eq:grid:aver}
\end{equation}
where $\SBnnft[\G{}]{}(\up,\as,\uff)$
\begin{equation}
  \Definition \!\!\dint{\upp}{\ass}{\!\!\!\!\gu{}(\up-\upp)\,\ga{}(\as-\ass)\,\SBnnft{}(\upp,\ass,\upp-\uff)}.
  \label{eq:grid:samp}
\end{equation}
We thus have equations that resemble those containing the sampling function
alone, except for 1) the replacement of the generalized sampling function
\SBnnft{} by its gridded version \SBnnft[\G{}]{} and 2) the way the
variables are linked together both in the gridding of \SBnnft{} (\ie{},
Eq.~\ref{eq:grid:samp}) and in the averaging of \SBftft[\G{}]{} (\ie{},
Eq.~\ref{eq:grid:aver}).

\subsection{Regular resampling}
\label{sec:resampling}

It is well known that too low a resampling rate in one space implies power
aliasing in the conjugate space~\citep[See \eg][]{bracewell00,press92}.
Aliasing must be avoided as much as possible because it folds power outside
the imaged region back into it. Table~\ref{tab:samplings} defines the
intervals of definition of the different functions we are dealing with
(\ie{}, visibilities, primary beam, dirty image, and dirty beam), as well
as the associated sampling rates needed to enforce Nyquist sampling. The
boundary values of the definition intervals ($\abs{\uf}_\emr{max}$ and
$\abs{\af}_\emr{max}$) are related to the sampling rates (\daf{} and
\duf{}, respectively) through
\begin{equation}
  \abs{\uf}_\emr{max}.\daf = \abs{\af}_\emr{max}.\duf = \frac{1}{\nsamp},
\end{equation}
where \nsamp{} is an integer characterizing the sampling. Nyquist sampling
implies $\nsamp=2$. However, slight oversampling (\eg{} $\nsamp=3$) is
often recommended because the measures suffer from errors and the
deconvolution is a nonlinear process. In this section, we examine the
properties of the different functions to define their associated sampling
rates.

\subsubsection{The \as{} sampling rate of the visibility function}

When Fourier transforming the measurement Eq.~\ref{eq:measurement:otf}
along the \as{} axis, we derive the Ekers \& Rots equation
(\ref{eq:ekers-rots:1}). This equation implies that \Vft{}(\up,\us) is
bounded inside the $[-\dprim,+\dprim]$ spatial frequency interval along the
\us{} axis. As a result, the visibility function needs to be regularly
resampled at a rate of only $0.5/\dprim$ to satisfy the Nyquist theorem.
This was first pointed out by~\citet{cornwell88}. This sampling rate is
equal to $\Aprim/2$ or $\about \Afwhm/2.4$.  The ``usual, wrong'' habit of
sampling at $\Afwhm/2$ is indeed undersampling with aliasing as a
consequence. \citet{mangum07} discuss the consequences of undersampling
in-depth in the framework of single-dish imaging.

\subsubsection{The \up{} sampling rate of the visibility function}

\TabSampling{} %
\FigPrimaryBeams{} %
\TabAliasing{} %

Now, the Fourier transform of the measurement
equation~\ref{eq:measurement:otf} along the \up{} axis gives
\begin{equation}
  \Vftnn{}(\ap,\as) = \B{}(\ap-\as)\,\I{}(\ap),
\end{equation}
\begin{equation}
  \mbox{where} \quad \FTop{\Vftnn{}(\ap,\as)}{\V{}(\up,\as)}{\ap}{\up}.
\end{equation}
We use the tilde sign under \V{} to denote the inverse Fourier transform of
\V{} along its first dimension.  A well-known Fourier transform property
implies that \B{} has infinite support because \Bft{} is bounded.  The
resampling rate along the \up{} axis therefore depends on the properties of
the product of $\B{}(\ap-\as)$ times $\I{}(\ap)$ as a function of \ap{}.
While no unique answer exists, three facts help us to find the right
sampling rate: 1) \B{} falls off relatively quickly. 2) The result depends
on the spatial distribution of the sky brightness and in particular on the
dynamic range in brightness needed to accurately image it. 3) The measure
of \Vftnn{}(\ap,\as) has a limited accuracy owing to thermal noise, phase
noise, and other possible systematics (\eg{} pointing errors). For
simplicity, we quantify the measurement accuracy by a single number, namely
the maximum instrumental fidelity measured in the image plane as defined
in~\citet{pety01}.  There are two cases:
\begin{enumerate}
\item The maximum instrumental fidelity limits the dynamic range in
  brightness. For instance, \citet{pety01} showed that the fidelity of
  interferometric imaging at (sub)-millimeter wavelengths will be limited
  to a few hundred. In this case, \Vftnn{}(\ap,\as) aliasing can be
  tolerated when the amplitude of \B{} is less than a fraction of the
  inverse of the maximum instrumental fidelity.
\item The maximum instrumental fidelity is much greater than the image
  fidelity, as can be the case at centimeter wavelengths. In this case,
  \Vftnn{}(\ap,\as) aliasing can only be tolerated when the amplitude of
  \B{} is less than a fraction of the inverse of the dynamic range of the
  image.
\end{enumerate}
The criterion derived in each case gives a typical image size (\Aalias),
which can be converted into the desired \up{} sampling rate. To be more
quantitative, Fig.~\ref{fig:primbeam} models the normalized antenna power
patterns of an antenna illuminated by a Gaussian beam of 12.5~dB edge taper
and with a given blockage factor (ratio of the secondary-to-primary
diameters). The top panel presents an ideal case without secondary miror,
while the middle and bottom panels present simple models of the ALMA and
PdBI antennas. The largest angular sizes at which the power patterns are
less than a given value, $\mathcal{P}_0$, is a first-order estimate of
$\Aalias/2$ to get a fidelity or dynamic range higher than
$1/\mathcal{P}_0$. Table~\ref{tab:aliasing} gives the values of
$\Aalias/\Afwhm$ as a function of the searched fidelity or dynamic range.
This condition is sufficient but not necessary. Indeed, the aliasing
properties also depend on the brightness distribution of the source.

\subsubsection{The \uf{} sampling rate of \Iwfft{}(\uf)}

We have no garantee that the sky outside the targeted field of view is
devoid of signal, so the only way to ensure a given dynamic range inside
the targeted field of view is to choose the image size large enough so that
the aliasing of potential outside sources is negligible. This means that
the dirty image size must be equal to the field-of-view size plus the
tolerable aliasing size
\begin{equation}
  \Aimage = \Afield + \Aalias. 
\end{equation}
The conjugate $uv$ distance and associated $uv$ sampling then are
\begin{equation}
  \dimage = \frac{\dfield}{1+\frac{\dfield}{\dalias}}
  \quad \mbox{and} \quad
  \duf = \frac{\dimage}{\nsamp}.
\end{equation}

\subsubsection{The \uff{} and \ufff{} sampling rates of \Dwftft{}(\uff,\ufff{})}

The \ufff{} axis must thus be sampled at the same rate as the second
dimension of the definition space of \Sft{}, \ie{}, as \us{}. Moreover,
\uff{} has in this equation a behavior ($\uff=\up+\usss$) similar to \uf{}
($=\up+\us$).  It must thus have the same sampling behavior as \uf{}. This
sampling rate ($\duff=\dimage/\nsamp$) is quite high. Some deconvolution
methods (see below) allow us to relax this sampling rate.

\subsection{Absence of gridding ``correction''}

Imaging of single-field observations goes through the following steps: 1)
convolution by a gridding kernel, 2) regular resampling, 3) fast Fourier
transform, and 4) gridding ``correction''. The so-called gridding
``correction'' is a division of the dirty beam and dirty image by the
Fourier transform of the gridding kernel used in the initial convolution.
This step is mandatory when imaging single-field observations to keep the
image-plane measurement equation as a simple convolution
equation~\citep[see \eg{}][]{sramek89}. When imaging wide-field
observations, as proposed here, the Fourier transform along the \as{}
dimension, followed by the shift-and-average operation, freeze the
convolution kernel into the dirty beam of the wide-field measurement
equation. This is why the gridding ``correction'' step is irrelevant here.

\section{Dirty beams, weighting, and deconvolution}
\label{sec:deconvolution}

In radioastronomy, the dirty beam is the response of the interferometer to
a point source. In the wide-field synthesis framework, the response of the
interferometer to a point source, \Dw{}, \emph{a priori} depends on the
source position on the sky. \Dw{}(\aff,\afff) can thus be interpreted as a
set of dirty beams, with each dirty beam referred to by its fixed \afff{}
sky coordinate. These simple facts raise several questions.  What are the
properties of the convolution kernel? Is it possible to modify these
properties? How do we deconvolve the dirty image?

\subsection{A set of wide-field dirty beams}

With the wide-field synthesis framework proposed here,
Appendix~\ref{sec:demo:dirty} shows that\\
$\Dw{}(\aff,\afff)=$
\begin{equation}
  \dint{\ap}{\as}{\!\!\!\!\!\!\!\B{}(\afff-\aff-\as)\Wa{}(\aff-\ap,\afff-\as)\Ds{}(\ap,\as)}.
  \label{eq:dirty-beam-4}
\end{equation}
\begin{equation}
  \mbox{where} \quad
  \FTop{\Ds{}(\ap,\as)}{\S{}(\up,\as)}{\ap}{\up},
\end{equation}
\begin{equation}
  \mbox{and} \quad
  \FTop{\Wa{}(\aff,\afff)}{\Wu{}(\uff,\ufff)}{(\aff,\afff)}{(\uff,\ufff)}.
  \label{eq:weight:ima}
\end{equation}
\Ds{}(\ap,\as) is the single-field dirty beam, associated with the $uv$
sampling at the sky coordinate \as{}. And \Wa{}(\aff,\afff) will be called
the image plane weighting function, while \Wu{}(\uff,\ufff) is the $uv$
plane weighting function. The set of wide-field dirty beams \Dw{} is then
the double convolution of the image plane weighting function and the
single-field dirty beams, apodized by the primary beam at the current sky
position \as{}.

While the shape of the single-field dirty beam is directly given by the
Fourier transform of the sampling function, the shape of the wide-field
dirty beam depends, directly or through Fourier transforms, on the sampling
function (\S{}), the primary beam shape (\B{}), and the weighting function
(\Wu{}). Moreover, the wide-field dirty beam shape \emph{a priori} varies
slowly with the sky position, since it is basically constant over the
primary beamwidth as stated in Sect.~\ref{sec:resampling}. It nevertheless
varies, implying, for instance, a ``slow'' variation of the synthesized
resolution over the whole field of view.

While the single-field and wide-field dirty beam expressions seem very
different, they share the same property of expressing the way the
interferometer is used to synthesize a telescope of larger diameter in the
image plane. In other words, the sampling function for single-field imaging
and \Dwftft{} for wide-field imaging express the sensitivity of the
interferometer to a given spatial frequency. These $uv$ space functions are
called the transfer functions of the interferometer~\citep[][chapter
5]{thompson86}. Modifying the transfer function has a direct impact on the
measured quantity. Once the interferometer is designed and the observations
are done, the only way to change this transfer function is data weighting.

An ideal set of wide-field dirty beams, \Dw{}(\aff,\afff), would have the
following properties. All the wide-field dirty beams should be identical
(\ie{}, independent of the \afff{} sky coordinate) and equal to a narrow
Gaussian (its FWHM giving the image resolution). \emph{This would give the
  product of a wide Gaussian of \uff{} by a Dirac function of \ufff{}, as
  the ideal wide-field transfer function, \Dwftft{}(\uff,\ufff).}

\subsection{Dirty beam shapes and weighting}

When imaging single-field observations, giving a multiplicative weight to
each visibility sample is an easy way to modify the shape of the dirty beam
and thus the properties of the dirty and deconvolved images.  Natural
weighting (which maximizes signal-to-noise ratio), robust weighting (which
maximizes resolution), and tapering (which enhances brightness sensitivity
at the cost of a lower resolution) are the most popular weighting
techniques~\citep[see \eg{}][]{sramek89}.

In the case of wide-field synthesis, a multiplicative weight can also be
attributed to each visibility sample before any processing. However, the
weighting is also at the heart of the wide-field synthesis because it is an
essential part of the shift-and-average operation. No constraint has been
set on the weighting function up to this point, which indicates that the
weighting function (\Wu{}) gives us a degree of freedom in the imaging
process. We look in turn at both kinds of weighting. In both cases, an
obvious issue is the definition of the optimum weighting functions. As in
the case of single-field imaging, there is no single answer to this
question. It depends on the conditions of the observation and on the
imaging goals.

\subsubsection{Weighting the measured visibilities}

Natural weighting consists of slightly changing the definition of the
sampling function. It is now set to a normalized natural weight where there
is a measure and 0 elsewhere. The natural weight is usually defined as the
inverse of the thermal noise variance, computed from the radiometric
equation, \ie{}, from the system temperature, the frequency resolution, and
the integration time. Using this weighting scheme before computing the
first Fourier transform along the \as{} sky dimension makes sense because
the observing conditions (and thus the noise) vary from visibility to
visibility.

We propose to generalize this weighting scheme to other observing
conditions than just the system noise. Indeed, critical limitations of
interferometric wide-field imaging are pointing errors, tracking errors,
atmospheric phase noise (in the (sub)-millimeter domain), etc. While
techniques exist for coping with these problems~\citep[\eg{}, water vapor
radiometer, direction-dependent gains:][]{bhatnagar08}, they are not
perfect. The usual way to deal with the remaining problems is to flag the
source data based on \emph{a priori} knowledge of the problems, \eg{},
pointing measurement, tracking errors, rms phase noise on calibrators, etc.
However, flagging involves the definition of thresholds, while reality is
never black and white. It can thus be asked whether some weighting scheme
could be devised to minimize the effect of pointing errors, tracking errors
or phase noise on the resulting image. We propose to modulate natural
weighting based on the \emph{a priori} knowledge of the observing
conditions.

\subsubsection{Weighting the synthesized visibilities}

Robust weighting or tapering the \emph{measured} visibilities do not make
sense in wide-field synthesis because the dirty image is made from the
synthesized visibilities after the first Fourier transform along the \as{}
sky dimension. A weighting function \Wu{} then appears naturally as part of
the shift-and-average operator. Its optimum value depends on the properties
of the measured sampling function. Here are a few examples.
\begin{description}
\item[\it{Infinite, continuous sampling.}] This is the ideal case studied
  in Sect.~\ref{sec:ideal}. Knowing that the Ekers \& Rots equation
  (\ref{eq:ekers-rots:1}) links the quantity we want to estimate, \ie{},
  \Ift{}, to many noisy\footnote{The noise is assumed to have a Gaussian
    probability distribution function.} measurements, \Vft{}(\up,\us{}),
  via a product by \Bft{} (assumed to be perfectly defined), we can invoke
  a simple least-squares argument~\citep[see \eg{}][]{bevington03} to
  demonstrate that the optimum weighting function is
  \begin{equation}
    \Wu{}(\up,\uf-\up) = \frac{\wu{}(\up,\uf-\up)\,\Bft{}(\up-\uf)}{\sint{\up}{\wu{}(\up,\uf-\up)\,\Bft{}^2(\up-\uf)}},
    \label{eq:weight}
  \end{equation}
  with \wu{}(\up,\us) the weight computed from the inverse of the noise
  variance of \Vft{}(\up,\us{}). Using Eq.~\ref{eq:dirty:uv}, it is then
  easy to demonstrate that $\Dwft{}(\uf) = 1$, and then $\Iwf{}(\af) =
  \I{}(\af)$. The dirty image is a direct estimate of the sky brightness;
  \ie{}, deconvolution is superfluous.
\item[\it{Complete sampling.}] The signal is Nyquist-sampled, but it has a
  finite support in both the $uv$ and sky planes, implying a finite
  synthesized resolution and a finite field of view. In contrast to the
  previous case, this one may have practical applications, \eg{},
  observations done with ALMA in its compact configuration. Indeed, the
  large number of independent baselines coupled to the design of the ALMA
  compact configuration ensure a complete, almost Gaussian, sampling for
  each snapshot. In this case, the best choice may be to choose the
  weighting function so that all the dirty beams are identical to the same
  Gaussian function. In this case, the deconvolution would also be
  superfluous.
\item[\it{Incomplete sampling.}] This is the more general case studied in
  Sect.~\ref{sec:real}. The signal not only has a finite support but it
  also is undersampled (at least in the $uv$ plane).  The deconvolution is
  mandatory. The choice of the weighting function thus will depend on
  imaging goals.
  
  If the user needs the best signal-to-noise ratio, some kind of natural
  weighting will be needed. It is tempting to use Eq.~\ref{eq:weight} as a
  natural weighting scheme. However, the main condition for derivation of
  this weighting function, \ie{}, the Ekers \& Rots equation
  (\ref{eq:ekers-rots:1}), is not valid anymore, as the noisy measured
  quantity (\SVft{}) is now linked to the quantity we want to estimate
  (\Ift{}) by a local average (see Eq.~\ref{eq:ekers-rots:3}).  This is why
  it was more appropriate to try to get a Gaussian dirty beam shape in the
  complete sampling case.
  
  If the signal-to-noise ratio is high enough, the user has two choices.
  Either he/she wants to maximize angular resolution power and needs some
  kind of robust weighting, or he/she wants to get the more homogeneous
  dirty beam shape over the whole field of view. This requirement cannot
  always be fully met.  The Ekers \& Rots scheme enables us to recover
  unmeasured spatial frequencies only in regions near to measured ones,
  because \Bft{} has a finite support.
\end{description}

\vspace{-\bigskipamount}

\subsection{Deconvolution}

Writing the image-plane measurement equation in a convolution-like way is
very interesting because all the deconvolution methods developed in the
past 30 years are optimized to treat deconvolution problems~\citep[see
\eg][]{hogbom74,clark80,schwab84,narayan86}. For instance, it should be
possible to deconvolve Eq.~\ref{eq:samp:meas:im} with just slight
modifications to the standard CLEAN algorithms. Indeed,
Eq.~\ref{eq:samp:meas:im} can be interpreted as the convolution of the sky
brightness by a set of dirty beams, so that the only change, once a CLEAN
component is found, would be the need to find the right dirty beam in this
set in order to remove the CLEAN component from the residual image.

Following~\citet{clark80} and ~\citet{schwab84}, most algorithms today
deconvolve in alternate minor and major cycles. During a minor cycle, a
solution of the deconvolution is sought with a simplified (hence
approximate) dirty beam. During a major cycle, the current solution is
subtracted either from the original dirty image using the exact dirty beam
or from the measured visibilities, implying a new gridding step. In both
cases, the major cycles result in greater accuracy. The iteration of minor
and major cycles enables one to find an accurate solution with better
computing efficiency. In our case, the approximate dirty beams used in the
minor cycle could be 1) dirty beams of a much smaller size than the image,
or 2) a reduced set of dirty beams (\ie{}, guessing that the typical
variation sizescale of the dirty beams with the sky coordinate is much
larger than the primary beamwidth), or 3) both simultaneously.  The model
would be subtracted from the original visibilities before re-imaging at
each major cycle. The trade-off is between the memory space needed to store
a full set of accurate dirty beams and the time needed to image the data at
each major step. Some quantitative analysis is needed to know how far the
dirty beams can be approximated in the minor cycle.

It is worth noting that the accuracy of the deconvolved image will be
affected by edge effects. Indeed, the dirty brightness at the edges of the
observed field of view is attenuated by the primary beam shape. When
deconvolving these edges, the deconvolved brightness will be less precise,
because the primary beam has a low amplitude there. This only affects the
edges, because inside the field of view, every sky position should be
observed a fraction of the time with a primary beam amplitude between 0.5
and 1.  This edge effect is nevertheless expected to be much less
troublesome than the inhomogeneous noise level resulting from standard
mosaicking imaging (see Sect.~\ref{sec:mosaicking:nutshell}).

\vspace{-\medskipamount}

\section{Short spacings}
\label{sec:short-spacings}

\subsection{The missing flux problem}

Radio interferometers are bandpass instruments; \ie{}, they filter out not
only the spacings longer than the largest baseline length but also the
spacings shorter than the shortest baseline length, which is typically
comparable to the diameter of the interferometer antennas. In particular,
radio interferometers do \emph{not} measure the visibility at the center of
the $uv$ plane (the so-called ``zero spacing''), which is the total flux of
the source in the measured field of view.

The lack of short baselines or \emph{short spacings} has strong effects as
soon as the size of the source is more than about 1/3 to 1/2 of the
interferometer primary beam.  Indeed, when the size of the source is small
compared to the primary beam of the interferometer, the deconvolution
algorithms use, in one way or another, the information of the flux at the
lowest \emph{measured} spatial frequencies for extrapolating the total flux
of the source.  The extreme case is a point source at the phase center for
which the amplitude of all the visibilities is constant and equal to the
total flux of the source: Extrapolation is then exact. However, the larger
the size of the source, the worse the extrapolation, which then
underestimates the total source flux. This is the well-known problem of the
missing flux that observers sometimes note when comparing a source flux
measured by a mm interferometer with the flux observed with a single-dish
antenna.

\FigPathLength{} %

Wide-field synthesis does \emph{not} recover the full short spacings.  Let
us assume that the visibility function is continuously sampled from $\dmin$
to $\dmax$, with $\dmin \about 1.5\,\dprim$. The length of the averaging
linepath\footnote{The notion of averaging linepath has been introduced in
  Sect.~\ref{sec:ideal} (see in particular Eq.~\ref{eq:average:1}).},
$L(\uf)$, can be interpreted as the number of measures that contribute to
the estimation of $\Ift{}(\uf)$.  Figure~\ref{fig:pathlength} shows the
variations of $L(\uf)$ function when starting from a visibility function
continuously defined in the $[\dmin,\dmax]$ interval along the \up{}
dimension. We can expect to recover $\Ift{}(\uf)$ only inside the
$[\dmin-\dprim,\dmax+\dprim]$ interval. In particular, information on short
spacings lower than $\dmin-\dprim$ (\eg{} the crucial zero spacing) cannot
be recovered when using a homegeneous interferometer, and the short
spacings in the interval $[\dmin-\dprim,\dmin]$ are recovered with
increasing accuracy from $\dmin-\dprim$ to $\dmin$. Both effects imply the
need for complementary instruments to accurately measure the missing
short-spacings.

\subsection{Usual hardware and software solutions}

To derive the correct result for larger source sizes, it is necessary to
complement the interferometer data with additional data, which contain the
missing short-spacing information. The IRAM-30m single-dish telescope is
used to complement the Plateau de Bure Interferometer. Short-spacing
information can also be in part recovered with a secondary array of smaller
antennas and shorter baselines (\eg{} the CARMA interferometer). In the
ALMA project, the short-spacing information will be derived by a
combination of four 12m-single-dish antennas and an interferometer of 12
antennas of 7 meters called ACA (Atacama Compact Array).

From the software point-of-view, two main families of algorithms exist in
the standard processing of mosaics. Either the short-spacing information is
combined on the deconvolved image (\ie{}, the interferometer data is imaged
and deconvolved separately) through a \emph{hybridization} in the Fourier
plane~\citep[see \eg{}][]{pety01}, or the long and short-spacing
information is imaged and/or deconvolved jointly. In this category, we find
the \emph{pseudo-visibility} technique, which produces interferometric-like
visibilities from single-dish maps~\citep[see \eg{}][and references
therein]{pety01,rodriguez08}, and the multi-resolution deconvolution
algorithms, which work on images containing different spatial frequency
ranges.

In the next two sections, we show how wide-field synthesis naturally
processes the short-spacing information either from single-dish or from
heterogeneous arrays.

\subsection{Processing short spacings from single-dish measurements}

The single-dish measurement equation can be written as
\begin{equation}
  \Isd{}(\af) = \Ssd{}(\af) \sint{\aff}{\Bsd{}(\aff-\af)\,\I{}(\aff)},
  \label{eq:sd:meas:1}
\end{equation}
where \Isd{} is the measured single-dish intensity, \Ssd{} the single-dish
sampling function, and \Bsd{} the single-dish antenna power pattern. As
already stated in the introduction, the above integral is identical to the
ideal measurement equation of interferometric wide-field imaging taken in
$\up=0$. If we define a single-dish visibility function as
\begin{equation}
  \Vsd{}(\up=0,\af) \Definition \sint{\aff}{\Bsd{}(\aff-\af)\,\I{}(\aff)},
\end{equation}
we can thus write the measured single-dish intensity as
\begin{equation}
  \Isd{}(\af) = \Ssd{}(\af)\,\Vsd{}(\up=0,\af).
  \label{eq:sd:meas:2}
\end{equation}
The recognition that the single-dish measurement equation is a particular
case of the interferometric wide-field measurement equation opens the way
to treating both the single-dish and interferometric data sets through
exactly the same processing steps. We just have to define a hybrid sampling
function, \Shyb{}, as
\begin{eqnarray}
  &&\Shyb{}(\up \neq 0,\af) = \S{}(\up,\af)\\
  &&\Shyb{}(\up = 0,\af) = \Ssd{}(\af),
\end{eqnarray}
the Fourier transform of the hybrid primary beam, \Bhybft{}, as
\begin{eqnarray}
  &&\Bhybft{}(\up \neq 0,\uff) = \Bft{}(\up-\uff)\\
  &&\Bhybft{}(\up = 0,\uff) = \Bsdft{}(-\uff),
\end{eqnarray}
and a hybrid weighting function, \Whyb{}, as
\begin{eqnarray}
  \!\!\!\!\!\!&&\Whyb{}(\up \neq 0,\uff+\ufff-\up) = \Whyb{}(\up,\uff+\ufff-\up),\\
  \!\!\!\!\!\!&&\Whyb{}(\up = 0,\uff+\ufff) = \Wsd{}(\uff+\ufff).
\end{eqnarray}
All the processing steps described in the previous sections (including a
potential gridding step of single-dish, on-the-fly data) can then be
directly applied to the hybrid data set. Using the wide-field synthesis
formalism, we can easily write
\begin{equation}
  \Ihybft{}(\uf) = \sint{\uff}{\Dhybftft{}(\uff,\uf-\uff)\,\Ift{}(\uff)},
\end{equation}
\begin{equation}
  \mbox{with} \quad
  \Ihybft{}(\uf) = \Iwfft{}(\uf) + \Wsd{}(\uf)\,\Isdft{}(\uf)
\end{equation}
and $\Dhybftft{}(\uff,\ufff)$
\begin{equation}
  = \Dwftft{}(\uff,\ufff) + \Wsd{}(\uff+\ufff)\,\Ssdft{}(\ufff)\,\Bsdft{}(-\uff).
\end{equation}
\TabSymbolShortSpacings{} %
We thus see that \Ihybft{} is a linear combination of the information
measured by the single-dish (\Isdft{}) and by the interferometer
(\Iwfft{}). There, \Wsd{}(\uf) plays a particular role for two reasons.
First, its dependency on the spatial frequency (\uf) enables us to filter
out the highest spatial frequencies that are measured by the single-dish
antenna with low accuracy. Second, it is well-known that the relative
weight of the single-dish to interferometric data is a critical parameter
in the processing of the short spacings from single-dish data~\citep[see
\eg{}][]{rodriguez08}. This relative weight is a free parameter within the
restrictions set by the noise level (\ie{}, we want the single-dish data to
bring information and not just noise to the interferometric data), and a
criterion must therefore be defined to adjust it to an optimal value. We
refer the reader to the discussion of Sect.~\ref{sec:deconvolution}, which
also applies here.

\subsection{Processing short spacings from heterogeneous arrays}

A heterogeneous array is an interferometer composed with antennas of
different diameters. ALMA and CARMA are two such examples. The measurement
equation for a heterogeneous array is
\begin{equation}
  \V{ij}(\up,\as) \!= \!\!\!\DFTint{\ap}{\!\!\!\!\!\Bv{i}(\ap-\as)\,\Bv[\star]{j}(\ap-\as)\,\I{}(\ap)}{\up},
  \label{eq:measurement:otf:het}
\end{equation}
where \Bv{i} and \Bv{j} are the voltage reception patterns of the antenna
pair that forms the $ij$ baseline and the asterisk denotes the complex
conjugate~\citep[][chapter~3]{thompson86}. The formalism developed in the
previous sections holds as long as we redefine
\begin{equation}
  \B{ij}(\af) \Definition \Bv{i}(\af)\,\Bv[\star]{j}(\af).
\end{equation}
A simple application of the correlation theorem implies that
\begin{equation}
  \Bft{ij}(\uf) = \sint{\uff}{\Bvft{i}(\uf+\uff)\,\Bvft{j}(\uff)}.
\end{equation}
The use of the baseline indices $ij$ must be generalized throughout the
equations because the knowledge of the antenna type must be attached to
each individual data point (visibility). As a result, the wide-field
synthesis formalism can be easily adapted to heterogeneous arrays at the
price of additional bookkeeping.

\subsection{Two textbook cases: IRAM-30m + PdBI and ALMA + ACA}

\FigTextbook{} %

Figure~\ref{fig:textbook} sketches why wide-field synthesis naturally
handles the short spacings in two textbook cases. In the ideal case, the
Fourier transform along the \as{} dimension produces visibilities, which
are related to the wide-field spatial frequencies of the source brightness
weighted by the transfer function of the interferometer. In this sense,
Fig.~\ref{fig:textbook} displays the natural weighting of the synthesized
wide-field visibilities at the position of each measured visibility.
Handling visibilities from antenna of different sizes just implies that the
natural weighting function of the synthesized visibilities will have a
different shape.

The top panel of Fig.~\ref{fig:textbook} displays how the IRAM-30m
single-dish is used to complement the Plateau de Bure interferometer
visibilities.  The bottom panel displays how ACA is used to produce the
short spacing information for ALMA. The four 12m-antennas will provide the
single-dish information, while the 12 additional 7m-antennas will form with
ALMA a heterogeneous array. In the first design, ACA and ALMA form two
independent interferometers; \ie{}, they are not cross-correlated.  The
single-dish antennas, ACA and ALMA, thus appear as three different
instruments.  It is thus possible to decompose the hybrid set of wide-field
dirty beams obtained by processing the 3 sets of data together in 3
different sets of dirty beams
\begin{equation}
  \Dhybftft{}(\uff,\ufff) = \Dwftft{12\emr{m}}(\uff,\ufff) + \Dwftft{7\emr{m}}(\uff,\ufff) + \Dsdftft{}(\uff,\ufff),
\end{equation}
with $\Dhybftft{}(\uff,\ufff)\Definition$
\begin{equation}
  \sint{\up}{\!\!\!\!\!\Whyb{}(\up,\uff+\ufff-\up)\,\Shybft{}(\up,\ufff)\,\Bhybft{}(\up,\uff)}.
  \label{eq:hybridbeamset}
\end{equation}
For a multiplying interferometer,
\begin{equation}
  \forall \abs{\up} < \dprim, \quad \S{}(\up,\as) = 0 
  \quad \mbox{and} \quad
  \Sft{}(\up,\us) = 0.
\end{equation}
This implies that \Dsdftft{} contributes at $\up = 0$ in the sum over \up{}
in Eq.~\ref{eq:hybridbeamset}, \Dwftft{7\emr{m}} contributes for
$9\,\emr{m} < \up \la 40\,\emr{m}$ and \Dwftft{12\emr{m}}(\uff,\ufff)
contributes for $15\,\emr{m} < \up < 150\,\emr{m}$ in the most compact
configuration of ALMA.

\section{Comparison with standard nonlinear mosaicking}
\label{sec:mosaicking}

\subsection{Mosaicking in a nutshell}
\label{sec:mosaicking:nutshell}

Several excellent descriptions of the mosaicking imaging and deconvolution
algorithms can be found~\citep[see \eg][]{cornwell88,cornwell93,sault96b}.
Here, we summarize the approach implemented in the \textsc{gildas/mapping}
software used to image and deconvolve the data from the Plateau de Bure
Interferometer. This approach is based on original ideas by F.~Viallefond
in the early 90s~\citep{gueth95}.

The basic ideas of nonlinear mosaicking are 1) imaging the different fields
of the mosaic independently, 2) linearly adding the single-field dirty
images into a dirty mosaic, and 3) jointly deconvolving the dirty mosaic.

\subsubsection{Single-field imaging} 

For simplicity, we skip the gridding convolution in the following equations
because the gridding step does not change the nature of the equations.
Imaging the fields individually means that we will work at constant \as{}.
We first define the single-field dirty image of the \as{}-field as
\begin{equation}
  \FTop{\Isf{}(\ap,\as)}{\Isfft{}(\up,\as)}{\ap}{\up},
\end{equation}
where the Fourier transform of the single-field dirty image is the product
of the sampling function $\S{}(\up,\as)$ and the visibility function
$\V{}(\up,\as)$:
\begin{equation}
  \Isfft{}(\up,\as) \Definition \SV{}(\up,\as).
\end{equation}
From the previous equations, it is easily demonstrated that
\begin{equation}
  \Isf{}(\ap,\as) = \!\!\sint{\app}{\!\!\!\!\Ds{}(\ap-\app,\as)\bracket{\B{}(\app-\as)\I{}(\app)}},
  \label{eq:jdirty}
\end{equation}
where the single-field dirty beam is defined as
\begin{equation}
  \FTop{\Ds{}(\ap,\as)}{\S{}(\up,\as)}{\ap}{\up}.
\end{equation}
We can rewrite the previous equation as
\begin{equation}
  \Isf[\as]{}(\ap) = \CONV{\Ds[\as]{}}{\paren{\B[\as]{}\I{}}}{\ap},
\end{equation}
meaning that the single-field dirty images can be written as a local
convolution of $\B[\as]{}\I{}$ and \Ds[\as]{}, the single-field dirty beam
associated to the currently imaged field.

\subsubsection{Mosaicking the dirty images}

\TabSymbolMosaicking{} %

In \textsc{gildas/mapping}\footnote{See
  \texttt{http://www.iram.fr/IRAMFR/GILDAS} for more information about the
  GILDAS software.}, the single-field dirty images are formed on the same
grid (in particular the same pixel size and the same image size covering
about twice the mosaic field of view). These single-field dirty images are
then linearly averaged as
\begin{equation}
  \Im{}(\ap) \Definition \sint{\as}{\Wm{}(\ap,\as)\,\Isf{}(\ap,\as)},
  \label{eq:jmosaic}
\end{equation}
\begin{equation}
  \mbox{where} \quad \Wm{}(\ap,\as) \Definition \frac{\wm{}(\as)\,\B{}(\ap-\as)}{\sint{\as}{\wm{}(\as)\,\B[2]{}(\ap-\as)}}
\end{equation}
and \wm{}(\as) is the sky plane weighting function, \ie{},
\begin{equation}
  \wm{}(\as) = \sum_i \dirac{}(\as-\ai) \frac{1}{\Nsf[2]{i}}.
\end{equation}
In the previous equation, the \ai{} are the positions of each sky-plane
measurement, and \Nsf{i} is the rms noise associated with \Isf[\ai]{}.
\citet{cornwell93} demonstrates that the noise in the mosaic image
naturally varies across the field as
\begin{equation}
  \Nm{}(\ap) = \frac{1}{\sqrt{\sint{\as}{\wm{}(\as)\,\B[2]{}(\ap-\as)}}}.
\end{equation}
In particular, it rises sharply at the edges of the mosaic.

\subsubsection{Joint deconvolution}

Standard algorithms of single-field deconvolution must be adapted to the
mosaicking case because both the dirty beam and the noise vary across the
mosaic field of view. We describe here the adaptations made in
\textsc{gildas/mapping} of the simplest CLEAN deconvolution method,
described in~\citet{hogbom74}. Adaptations of more evolved CLEAN
deconvolution methods are also implemented following the same basic rules.
\begin{enumerate}
\item We first initialize the residual and signal-to-noise maps from the
  dirty and noise maps
  \begin{equation}
    \Rm{0}(\ap) = \Im{}(\ap)
  \end{equation}
  \begin{equation}
    \mbox{and} \quad \SNRm{0}(\ap) = \frac{\Im{}(\ap)}{\Nm{}(\ap)}.
  \end{equation}
\item The $k^\emr{th}$ CLEAN component is sought on the \SNRm{k-1} map
  instead of the \Rm{k-1} map to ensure that noise peaks at the edges of
  the mosaic are not confused with the true signal of the same magnitude.
\item Using that the $k^\emr{th}$ CLEAN component is a point source of
  intensity \I{k} at position \ak{}, the residual and
  signal-to-noise maps are then upgraded as follows:\\
  $\Rm{k}(\ap) = \Rm{k-1}(\ap)$
  \begin{equation}
    -\gamma\I{k}\!\!\sint{\as}{\!\!\!\!\Wm{}(\ap,\as)\,\Ds{}(\ap-\ak,\as)\,\B{}(\ak-\as)},
  \end{equation}
  \begin{equation}
    \mbox{and} \quad \SNRm{k}(\ap) = \frac{\Rm{k}(\ap{})}{\Nm{}(\ap)}.
  \end{equation}
  Here $\gamma (\sim 0.2)$ is the usual loop gain that ensures convergence
  of the CLEAN algorithms.
\item Steps 2 and 3 are iterated as long as the stopping criterion is not
  met.
\end{enumerate}

\subsubsection{Wide-field measurement equation}

To help the comparison between mosaicking and wide-field synthesis, we now
go one step further than is usually done in the description of mosaicking;
\ie{}, we write the image-plane measurement equation as a wide-field
measurement equation of the same kind as Eq.~\ref{eq:samp:meas:im}.
Substituting Eq.~\ref{eq:jdirty} into Eq.~\ref{eq:jmosaic} and reordering
the terms after inverting the order of the sum over \as{} and \ap{}, one
obtains
\begin{equation}
  \Im{}(\ap) = \sint{\app}{\Dm{}(\ap-\app,\ap)\,\I{}(\app)},  
   \label{eq:mos-meas-eq:im}
\end{equation}
with $\Dm{}(\aff,\afff)$
\begin{equation}
   = \sint{\as}{\B{}(\afff-\aff-\as)\,\Wm{}(\afff,\as)\,\Ds{}(\aff,\as)}.
   \label{eq:mos-dirty-beam:im}
\end{equation}
Taking the inverse Fourier transforms of \Dm{}, we get the mosaicking
transfer function\\
$\Dmftft{}(\uff,\uf-\uff)$=
\begin{equation}
  \dint{\up}{\us}{\!\!\!\!\!\Wmu{}(\uf-\up,\us-\uff)\,\Sft{}(\up,\up-\us)\,\Bft{}(\up-\uff)},
  \label{eq:mos-dirty-beam:uv}
\end{equation}
\begin{equation}
  \mbox{with} \quad
  \FTop{\Wm{}(\aff,\afff)}{\Wmu{}(\uff,\ufff)}{(\aff,\afff)}{(\uff,\ufff)}.
\end{equation}

\subsection{Comparison}

While both mosaicking and wide-field synthesis produce image-plane
measurement equations of the same kind (see Eqs.~\ref{eq:samp:meas:im}
and~\ref{eq:mos-meas-eq:im}), the comparison of the dirty beams
(Eqs.~\ref{eq:dirty-beam-4} and~\ref{eq:mos-dirty-beam:im}) and of the
transfer functions (Eqs.~\ref{eq:dirty-beam}
and~\ref{eq:mos-dirty-beam:uv}) immediately shows the different
dependencies on the primary beams (\B{}), the single-field dirty beams
(\Ds{}), the image-plane weighting functions (\Wa{}), and their respective
Fourier transforms (\Bft{}, \Sft{} and \Wu{}). This means that mosaicking
is not mathematically equivalent to wide-field synthesis, though both
methods recover the sky brightness. These differences come directly from
the differences in the processing.  If we momentarily forget the gridding
steps, mosaicking starts with a Fourier transform along the \up{} dimension
of the visibility function, and most of the processing thus happens in the
sky plane, while wide-field synthesis starts with a Fourier transform along
the \as{} dimension, and most of the processing thus happens in the $uv$
plane.

Moreover, both methods are irreducible to each other. Wide-field synthesis
gives a more complex dirty beam formulation in the image plane, which could
give the impression that it is a generalization of mosaicking.  Indeed, the
wide-field image-plane weighting function can be chosen as the product of a
Dirac function of \aff{} times a function \Waa{} of \afff{},
\begin{equation}
  \mbox{\ie{},} \quad \Wa{}(\aff,\afff) = \dirac(\aff) \, \Waa{}(\afff).
\end{equation}
This implies a wide-field $uv$-plane weighting function independent of
\uff{}; \ie{}, \Wu{}(\uff,\ufff) = \Waaft{}(\ufff). This choice is a clear
limitation because it enables us to influence the transfer function only
locally (around each measured \up{} spatial frequency), while weighting is
generally intended to globally influence the transfer function (see
Sect.~\ref{sec:deconvolution}). Eitherway, in this case, the wide-field
dirty beam can easily be simplified to
\begin{equation}
  \Dw{}(\aff,\afff) \!= \!\!\!\sint{\as}{\!\!\!\!\B{}(\afff-\aff-\as)\,\Waa{}(\afff-\as)\,\Ds{}(\aff,\as)}.
\end{equation}
While this simplified formulation of the wide-field dirty beam is closer to
the mosaicking formulation, they still differ in a major way:
\Waa{}(\afff-\as) is a shift-invariant function contrary to
\Wm{}(\afff,\as). This is the shift-dependent property of \Wm{}(\afff,\as),
which implies the additional complexity (integral over \us{} in addition to
the integral over \up{}) of the mosaicking (Eq.~\ref{eq:mos-dirty-beam:uv})
over the wide-field transfer function (Eq.~\ref{eq:dirty-beam}).

One main difference between the two processing methods is that standard
mosaicking prescribes a precise weighting function, while we argue that the
wide-field weighting function should be defined according to the context
(see Sect.~\ref{sec:deconvolution}). Another important difference is the
treatment of the short spacings, which are naturally processed in the
wide-field synthesis methods, but which needs a very specific treatment in
mosaicking (see Sect.~\ref{sec:short-spacings} and references therein).
Finally, while mosaicking implies a gridding only of \up{} dimension of the
measured visibilities, wide-field synthesis naturally requires a gridding
of both the \up{} and \as{} dimensions. As the Nyquist sampling along the
\as{} dimension is only $0.5/\dprim$, the gridding of the sky plane can
result in a large reduction of the data storage space and cpu processing
cost when processing on-the-fly and/or multi-beam observations.

\section{Summary}

Interferometric wide-field imaging implies scanning the sky in one way or
another (\eg{} stop-and-go mosaicking, on-the-fly scanning, sampling of the
focal plane by multi-beams). This produces sampled visibilities \SV{},
which depends both on the $uv$-plane and sky coordinates (\eg{}, \up{} and
\as{}).

Based on a basic idea by~\citet{ekers79}, we proposed a new way to image
the interferometric wide-field sampled visibilities: \SV{}(\up,\as).  After
gridding the measured visibilities both in the $uv$ and sky planes, the
gridded visibilities \SV[\G{}]{} are Fourier-transformed along the \as{}
sky dimension, yielding synthesized visibilities \SVft[\G{}]{} sampled on a
$uv$ grid whose cell size is related to the total field of view; \ie{}, it
is much finer than the diameter of the interferometer antennas. We thus
proposed calling this processing scheme ``wide-field synthesis''.

The Fourier transform is performed for each constant \up{} value. As many
independent estimates of the $uv$ plane are produced as independent values
of \up{} measured.  A shift-and-average operator is then used to build a
final, wide-field $uv$ plane, which translates into a wide-field dirty
image after inverse Fourier transform, \ie{},
\begin{equation}
  \Iwfft[\G{}](\uf) \Definition \sint{\up}{\Wu{}(\up,\uf-\up)\,\SVft[\G{}]{}(\up,\uf-\up)},
  \label{eq:shift-and-average}
\end{equation}
where \Wu{} is a normalized weighting function. Using these tools, we
demonstrated that:
\begin{enumerate}
\item The dirty image (\Iwf[\G{}]{}) is a convolution of the sky brightness
  distribution (\I{}) with a set of wide-field dirty beams (\Dw[\G{}]{})
  varying with the sky coordinate \af{}, \ie{},
  \begin{equation}
    \Iwf[\G{}](\af) = \sint{\aff}{\Dw[\G{}]{}(\af-\aff,\af)\,\I{}(\aff)}.
  \end{equation}
  Compared to single-field imaging, the dependency on the primary beam is
  transferred from a product of the sky brightness distribution into the
  definition of the set of wide-field dirty beams.
\item The set of gridded dirty beams (\Dw[\G{}]{}) can be computed from the
  ungridded sampling function (\S{}), the transfer function (\Bft{}, the
  inverse Fourier transform of the primary beam), and the gridding
  convolution kernel (See Eq.~\ref{eq:samp:def:1}, \ref{eq:grid:aver}
  and~\ref{eq:grid:samp}).
\item The dependence of the wide-field dirty beams on the sky position is
  slowly-varying, with their shape varying on an angular scale typically
  larger than or equal to the primary beamwidth.
\end{enumerate}
Adaptations of the existing deconvolution algorithms should be
straightforward.

A comparison with standard nonlinear mosaicking shows that it is not
mathematically equivalent to the wide-field synthesis proposed here, though
both methods do recover the sky brightness. The main advantages of
wide-field synthesis over standard nonlinear mosaicking are
\begin{enumerate}
\item Weighting is at the heart of the wide-field synthesis because it is
  an essential part of the shift-and-average operation. Indeed, not only
  can a multiplicative weight be attributed to each visibility sample
  before any processing, but the $uv$-plane weighting function (\Wu{}, see
  Eq.~\ref{eq:shift-and-average}) is also a degree of freedom, which should
  be set according to the conditions of the observation and the imaging
  goals, \eg{} highest signal-to-noise ratio, highest resolution, or most
  uniform resolution over the field of view.  The \Wu{} weighting function
  thus enables us to modify the wide-field response of the instrument. On
  the other hand, mosaicking requires a precise weighting function in the
  image plane, which freezes the wide-field response of the interferometer.
\item Wide-field synthesis naturally processes the short spacings from both
  single-dish antennas and heterogeneous arrays along with the long
  spacings. Both of them can then be jointly deconvolved.
\item The gridding of the sky plane dimension of the measured visibilities,
  required by the wide-field synthesis, may potentially save large amounts
  of hard-disk space and cpu processing power relative to mosaicking when
  handling data sets acquired with the on-the-fly observing mode.
  Wide-field synthesis could thus be particularly well suited to process
  on-the-fly observations.
\end{enumerate}

The wide-field synthesis algorithm is compatible with the $uvw$-unfaceting
technique devised by~\citet{sault96a} to deal with the celestial projection
effect, known as non-coplanar baselines (see
appendix~\ref{sec:projection}). Finally, on-the-fly observations imply an
elongation of the primary beam along the scanning direction. These effects
can be decreased by an increase in the primary beam sampling rate.
However, it may limit the dynamic range of the image brightness if the
primary beam sampling rate is too coarse (see appendix~\ref{sec:otf}).

\begin{acknowledgements}
  This work has mainly been funded by the European FP6 ``ALMA enhancement''
  grant. This work was also funded by grant ANR-09-BLAN-0231-01 from the
  French {\it Agence Nationale de la Recherche} as part of the SCHISM
  project. The authors thank F.~Gueth for the management of the on-the-fly
  working package of the ``ALMA enhancement'' project.  They also thank
  S.~Guilloteau, R.~Lucas and J.~Uson for useful comments at various stages
  of the manuscript and D.~Downes for editing their English.  They finally
  thank the referee, B.~Sault, for his insightful comments, which
  challenged us to try to write a better paper.
\end{acknowledgements}
 
\bibliographystyle{aa} %
\bibliography{ms12873} %

\newpage{}

\appendix{}

\section{Demonstrations}
\label{sec:demo}

\subsection{Ekers \& Rots scheme}
\label{sec:demo:er}

Fourier-transforming the visibility function along the \as{} dimension at
constant \up{}, we derive with simple replacements
\begin{eqnarray}
  && \Vft{\up}(\us) \\
  &=& \DFTint{\as}{\V{\up}(\as)}{\us} \\
  &=& \dint{\as}{\ap}{\B{}(\ap-\as)\,\I{}(\ap)\,\dexp{(\ap\up+\as\us)}}.
\end{eqnarray}
We then use the following change of variables $\ab \Definition \ap-\as$ and
$d\ab = -d\as$, to get
\begin{eqnarray}
  \!\!\! && \Vft{\up}(\us) \\
  \!\!\! &=& \dint{\ap}{\ab}{\B{}(\ab)\,\I{}(\ap)\,\dexp{\bracket{\ap(\up+\us)-\ab\us}}} \\
  \!\!\! &=& \!\!\!\bracket{\!\DFTint{\ab}{\!\!\B{}(\ab)}{(-\us)}}\!\!\!\bracket{\!\DFTint{\ap}{\!\!\!\!\I{}(\ap)}{(\up+\us)}} \\
  \!\!\! &=& \Bft{}(-\us)\,\Ift{}(\up+\us).
\end{eqnarray}

\subsection{Incomplete sampling}
\label{sec:demo:samp}

We here demonstrate that Eq.~\ref{eq:samp:meas:im} and
Eq.~\ref{eq:samp:meas:uv} are equivalent. To do this, we take the direct
Fourier transform of $\Iwf(\af)$
\begin{eqnarray}
  && \Iwfft{}(\uf) \\
  &=& \dint{\af}{\aff}{\Dw{}(\af-\aff,\af)\,\I{}(\aff)\dexp{\af\uf}},
\end{eqnarray}
and we replace $\I{}(\aff)$ by its formulation as a function of its Fourier
transform
\begin{equation}
  \I{}(\aff) = \IFTint{\uff}{\Ift{}(\uff)}{\aff}.
\end{equation}
We thus derive $\Iwfft{}(\uf)$
\begin{equation}
  = \!\!\sint{\uff}{\!\bracket{\dint{\af}{\aff}{\!\!\!\!\Dw{}(\af-\aff,\af)\dexp{(\af\uf-\aff\uff)}}}\!\Ift{}(\uff)}.
\end{equation}
Using the following change of variables $\afff \Definition \af-\aff$, $\aff
= \af-\afff$ and $d\afff = -d\aff$, the innermost integral can be written
as
\begin{eqnarray}
  && \dint{\af}{\aff}{\Dw{}(\af-\aff,\af)\,\dexp{(\af\uf-\aff\uff)}}\\
  &=& \DFTint{\af}{\bracket{\DFTint{\afff}{\!\!\!\!\Dw{}(\afff,\af)}{\uff}}}{(\uf-\uff)}\\
  &=& \DFTint{\af}{\Dwftnn{}(\uff,\af)}{(\uf-\uff)}\\
  &=& \Dwftft{}(\uff,\uf-\uff).
\end{eqnarray}
In the last two steps, we have simply recognized two different steps of
Fourier transforms of \Dw{}. Finally,
\begin{equation}
  \Iwfft{}(\uf) = \sint{\uff}{\Dwftft{}(\uff,\uf-\uff)\,\Ift{}(\uff)}.
\end{equation}

\subsection{Gridding}
\label{sec:demo:grid}

The gridding kernel can be defined as the product of two functions, each
one operating in its own dimension. We use this to study separately the
effect of gridding in the $uv$ and sky planes. We then use the intermediate
results to get the effect of gridding simultaneously in both planes.

\subsection{Gridding in the $uv$ plane}
\label{sec:demo:grid:uv}

We define the sampled visibility function gridded in the $uv$ plane as
\begin{eqnarray}
  && \SV[\gu{}]{}(\up,\as) \Definition \CONV{\gu{}}{\SV[\as]{}}{\up}\\
  &=& \conv{\gu{}}{\SV[\as]{}}{\upp}{\up},
\end{eqnarray}
Using that the gridding is here applied on the \up{} dimension, while the
Fourier transform is applied on the \as{} dimension, it is easy to show
that the gridding and Fourier-transform operations commute:
\begin{eqnarray}
  && \SVft[\gu{}]{\up}(\us) \\
  &=& \!\!\!\dint{\as}{\upp}{\!\!\!\!\!\gu{}(\up-\upp)\S{}(\upp,\as)\V{}(\upp,\as)\dexp{\as\us}}\\
  &=& \sint{\upp}{\gu{}(\up-\upp)\,\SVft{\upp}(\us)}
\end{eqnarray}
Defining the Fourier transform of the $uv$ gridded dirty image, we derive
\begin{eqnarray}
  \!\!&& \Iwfft[\gu{}](\uf) \Definition \average{\SVft[\gu{}]{}}(\uf)\\
  \!\!&=& \!\!\!\dint{\up}{\upp}{\!\!\!\!\!\Wu{}(\up,\uf-\up)\,\gu{}(\up-\upp)\,\SVft{\upp}(\uf-\up)}.
\end{eqnarray}
Using Eq.~\ref{eq:ekers-rots:3} to replace $\SVft{\upp}(\uf-\up)$, we can
write the Fourier transform of the $uv$ gridded dirty image as
\begin{equation}
  \Iwfft[\gu{}](\uf) = \sint{\uff}{\Dwftft[\gu{}]{}(\uff,\uf-\uff)\,\Ift{}(\uff)},
\end{equation}
with
\begin{equation}
   \Dwftft[\gu{}]{}(\uff,\uf-\uff) \Definition \!\!\sint{\up}{\!\!\!\!\Wu{}(\up,\uf-\up)\,\SBftft[\gu{}]{}(\up,\uf-\up,\uff)}
\end{equation}
and $\SBftft[\gu{}]{}(\up,\us,\uff)$
\begin{equation}
  \Definition \sint{\upp}{\!\!\!\!\gu{}(\up-\upp)\,\Sft{}(\upp,\us-\uff+\upp)\,\Bft{}(\upp-\uff)}.
\end{equation}
Using $\Sft{}(\upp,\us-\uff+\upp)$
\begin{equation}
  = \bracket{\DFTint{\as}{\S{\upp}(\as)}{\us}} \dexp{(\upp-\uff)\as},
\end{equation}
\begin{equation}
  \mbox{and} \quad
  \FTop{\SBnnft[\gu{}]{}(\up,\as,\uff)}{\SBftft[\gu{}]{}(\up,\us,\uff)}{\as}{\us},
\end{equation}
we derive $\SBnnft[\gu{}]{}(\up,\as,\uff) =$
\begin{equation}
  \sint{\upp}{\!\!\!\!\gu{}(\up-\upp)\,\S{}(\upp,\as)\,\Bft{}(\upp-\uff)\,\dexp{(\upp-\uff)\as}\!},
\end{equation}
or
\begin{equation}
  \SBnnft[\gu{}]{}(\up,\as,\uff) = \!\!\sint{\upp}{\!\!\!\!\gu{}(\up-\upp)\,\SBnnft{}(\upp,\as,\upp-\uff)}.
  \label{eq:samp:grid:uv}
\end{equation}
Thus, \SBnnft[\gu{}]{} is the $uv$ gridded version of the generalized
sampling function \SBnnft{}.

\subsubsection{Gridding in the sky plane}
\label{sec:demo:grid:sky}

We define the sampled visibility function gridded in the sky plane as
\begin{eqnarray}
  && \SV[\ga{}]{}(\up,\as) \Definition \CONV{\ga{}}{\SV{\up}}{\as}\\
  &=& \conv{\ga{}}{\SV{\up}}{\ass}{\as},
\end{eqnarray}
Applying the convolution theorem on the Fourier transform along the \as{}
dimension, we derive
\begin{equation}
  \SVft[\ga{}]{\up}(\us) = \gaft{}(\us)\,\SVft{\up}(\us).
  \label{eq:grid:sky}
\end{equation}
Defining the Fourier transform of the sky-gridded dirty image, we derive
\begin{eqnarray}
  && \Iwfft[\ga{}](\uf) \Definition \average{\SVft[\ga{}]{}}(\uf)\\
  &=& \sint{\up}{\Wu{}(\up,\uf-\up)\,\gaft{}(\uf-\up)\,\SVft{\up}(\uf-\up)}.
\end{eqnarray}
Using Eq.~\ref{eq:ekers-rots:3} to replace $\SVft{\up}(\uf-\up)$, we can
write the Fourier transform of the sky-gridded dirty image as
\begin{equation}
  \Iwfft[\ga{}](\uf) = \sint{\uff}{\Dwftft[\ga{}]{}(\uff,\uf-\uff)\,\Ift{}(\uff)}  
\end{equation}
with $\Dwftft[\ga{}]{}(\uff,\uf-\uff)$
\begin{equation}
  \Definition \!\!\sint{\up}{\!\!\!\!\Wu{}(\up,\uf-\up)\,\SBftft[\ga{}]{}(\up,\uf-\up,\up-\uff)}
\end{equation}
\begin{equation}
  \mbox{and} \quad \SBftft[\ga{}]{}(\up,\us,\usss) \Definition \gaft{}(\us)\,\Sft{}(\up,\us+\usss)\,\Bft{}(\usss),
\end{equation}
or, with the definition of \SBftft{} (\ie{}, Eq.~\ref{eq:samp:def:2}),
\begin{equation}
  \SBftft[\ga{}]{}(\up,\us,\usss) \Definition \gaft{}(\us)\,\SBftft{}(\up,\us,\usss).
\end{equation}
\begin{equation}
  \mbox{Using} \quad \FTop{\SBnnft[\ga{}]{}(\up,\as,\usss)}{\SBftft[\ga{}]{}(\up,\us,\usss)}{\as}{\us},
\end{equation}
and the convolution theorem when taking the inverse Fourier transform of
\SBftft[\ga{}]{}, we derive
\begin{equation}
  \SBnnft[\ga{}]{}(\up,\as,\usss) = \sint{\ass}{\ga{}(\as-\ass)\,\SBnnft{}(\up,\ass,\usss)}.
\end{equation}
Thus, \SBnnft[\ga{}]{} is the sky gridded version of the generalized
sampling function \SBnnft{}.

\subsection{Gridding in both planes}
\label{sec:demo:grid:both}

Starting from the definition of \SV[\G{}]{} (Eq.~\ref{eq:grid:conv}), we
Fourier-transform it along the sky dimension at constant \up{}. Using that
the gridding along the \up{} dimension can be factored out of the Fourier
transform, we derive
\begin{equation}
  \SVft[\G{}]{\up}(\us) = \sint{\upp}{\gu{}(\up-\upp)\,\SVft[\ga{}]{\upp}(\us)}.
\end{equation}
Using Eq.~\ref{eq:grid:sky}, we now replace \SVft[\ga{}]{\upp}(\us) in the
previous equation to get
\begin{equation}
  \SVft[\G{}]{\up}(\us) = \gaft{}(\us) \sint{\upp}{\gu{}(\up-\upp)\,\SVft{\upp}(\us)},
\end{equation}
\begin{equation}
  \mbox{or} \quad \SVft[\G{}]{\up}(\us) = \gaft{}(\us) \SVft[\gu{}]{\up}(\us).
\end{equation}
From this relation, it is easy to deduce that
\begin{equation}
  \SBftft[\G{}]{}(\up,\us,\uff) = \gaft{}(\us) \SBftft[\gu{}]{}(\up,\us,\uff).
\end{equation}
Using the convolution theorem when taking the inverse Fourier transform of
\SBftft[\G{}]{} along the \us{} dimension and replacing
\SBnnft[\gu{}]{}(\up,\ass,\uff) with Eq.~\ref{eq:samp:grid:uv}, we finally derive\\
$\SBnnft[\G{}]{}(\up,\as,\uff)$
\begin{equation}
  \Definition \!\!\!\dint{\upp}{\ass}{\!\!\!\!\!\gu{}(\up-\upp)\,\ga{}(\as-\ass)\,\SBnnft{}(\upp,\ass,\upp-\uff)}.
\end{equation}

\subsection{Wide-field vs single-field dirty beams}
\label{sec:demo:dirty}

\newcommand{\intone}{\emm{\emr{FT}_1(\up,\uff,\afff)}}
\newcommand{\inttwo}{\emm{\emr{FT}_2(\up,\as,\aff,\afff)}}
\newcommand{\intthr}{\emm{\emr{FT}_3(\as,\aff,\afff)}}

The notation~\ref{eq:weight:ima} yields $\Wu{}(\uff,\ufff) =
\Waftft{}(\uff,\ufff)$. Using this in Eq.~\ref{eq:dirty-beam} gives\\
$\Dwftft{}(\uff,\ufff)$
\begin{equation}
  = \sint{\up}{\!\!\!\Waftft{}(\up,\uff+\ufff-\up)\,\Sft{}(\up,\ufff)\,\Bft{}(\up-\uff)}.
  \label{eq:dirty-beam-3}
\end{equation}
Taking the inverse Fourier transform along the \ufff{} axis of
Eq.~\ref{eq:dirty-beam-3} and reordering the integral to factor out the
term independent of \ufff{}, we can write
\begin{equation}
  \Dwftnn{}(\uff,\afff) = \sint{\up}{\!\!\!\Bft{}(\up-\uff)\,\intone},
  \label{eq:demo:dirty:1}
\end{equation}
where
\begin{eqnarray}
  \!\!\! &           & \!\!\! \intone \\
  \!\!\! &\Definition& \!\!\! \IFTint{\ufff}{\!\Waftft{}(\up,\uff+\ufff-\up)\,\Sft{}(\up,\ufff)}{\afff}.
\end{eqnarray}
We now introduce the following definition
\begin{equation}
  \Sft{}(\up,\ufff) \Definition \DFTint{\as}{\!\!\!\S{}(\up,\as)}{\ufff},
\end{equation}
to derive
\begin{eqnarray*}
  \!\!\! & & \!\!\! \intone \\
  \!\!\! &=& \!\!\! \sint{\as}{\!\!\!\!\!\S{}(\up,\as)\!\bracket{\!\sint{\ufff}{\!\!\!\!\!\Waftft{}(\up,\uff+\ufff-\up)\iexp{\ufff(\afff-\as)}}\!}\!\!}.
\end{eqnarray*}
Using the following change of variables $v \Definition \ufff+\uff-\up$,
$\ufff = v-\uff+\up$ and $dv = d\ufff$ on the innermost integral, we get
\begin{eqnarray*}
  \!\!\! & & \!\!\! \intone \\
  \!\!\! &=& \!\!\! \sint{\as}{\!\!\!\S{}(\up,\as)\,\Waftnn{}(\up,\afff-\as)\,\iexp{(\up-\uff)(\afff-\as)}}.
\end{eqnarray*}
Substituting this result into Eq.~\ref{eq:demo:dirty:1} and taking the
inverse Fourier transform along the \uff{} axis, we can write\\
$\Dw{}(\aff,\afff) =$
\begin{equation}
  \dint{\up}{\as}{\!\!\!\!\!\!\Waftnn{}(\up,\afff-\as)\,\S{}(\up,\as)\,\inttwo},
  \label{eq:demo:dirty:2}
\end{equation}
where
\begin{eqnarray*}
  \!\!\! &           & \!\!\! \inttwo\\
  \!\!\! &\Definition& \!\!\! \sint{\uff}{\Bft{}(\up-\uff)\,\iexp{\up(\afff-\as)}\,\iexp{\uff(\aff-\afff+\as)}}.
\end{eqnarray*}
Using the following change of variables $v \Definition \up-\uff$, $\uff =
\up-v$ and $dv = d\uff$, we get
\begin{equation}
  \inttwo = \B{}(\afff-\aff-\as)\,\iexp{\up\aff}.
\end{equation}
Substituting this result into Eq.~\ref{eq:demo:dirty:2} and re-ordering the
terms, we can write
\begin{equation}
  D{}(\aff,\afff) = \sint{\as}{\!\!\!B{}(\afff-\aff-\as)\,\intthr},
  \label{eq:demo:dirty:3}
\end{equation}
where
\begin{eqnarray*}
  \intthr \!\!&\Definition&\!\!\! \sint{\up}{\!\!\!\Waftnn{}(\up,\afff-\as)\,\S{}(\up,\as)\,\iexp{\up\aff}}.
\end{eqnarray*}
A simple application of the convolution theorem gives
\begin{eqnarray*}
 \intthr \!\!&\Definition&\!\!\! \sint{\ap}{\!\!\!\Wa{}(\aff-\ap,\afff-\as)\,\Ds{}(\ap,\as)},
\end{eqnarray*}
\begin{equation}
  \mbox{where} \quad \FTop{\Ds{}(\ap,\as)}{\S{}(\up,\as)}{\ap}{\up}.
\end{equation}
Substituting this result into Eq.~\ref{eq:demo:dirty:3}, we finally derive
the desired expression, \ie{}, Eq.~\ref{eq:dirty-beam-4}.

\section{From the celestial sphere onto a single tangent plane}
\label{sec:projection}

Eq.~\ref{eq:measurement:otf} neglects projection effects, known as
non-coplanar baselines. Any method which deals with interferometric
wide-field imaging must take this problem into account. After a short
introduction to the problem, we show how wide-field synthesis is compatible
with at least one method, namely the $uvw$-unfaceting of \citet{sault96b}.
This method tries to build a final wide-field $uv$ plane from different
pieces, just as our wide-field synthesis approach does. Another promising
method is the $w$-projection, based on original ideas of~\citet{frater80}
and first successfully implemented by~\citet{cornwell08}. We did not look
yet at its compatibility with wide-field synthesis.

\subsection{$w$-axis distortion}

When projection effects are taken into account, the measurement equation reads\\
$\V{}(\wp,\up,\as)$
\begin{equation}
  =\!\!\!\sint{\ap}{\!\!\!\!\!\B{}(\ap-\as)\,\frac{\I{}(\ap)}{\sqrt{1-\ap^2}}\,\dexp{\sbracket{\ap\up+\wp(\sqrt{1-\ap^2}-1)}}}.
  \label{eq:measurement:otf:3}
\end{equation}
In this equation, we continue to work in 1 dimension for the sky cosine
direction (\ap{}), but we explicitly introduce the dependence along the
direction perpendicular to the sky plane. This dependence appears in two
ways, which is handled in very different ways. First, the factor
$\sqrt{1-\ap^2}$ can be absorbed into a generalized sky brightness function
\begin{equation}
  \Ig{}(\ap) \Definition \frac{\I{}(\ap)}{\sqrt{1-\ap^2}}.
\end{equation}
After imaging and deconvolution, \I{}(\ap) can be easily restored from the
deconvolved \Ig{}(\ap) image. The second dependence appears as an
additional phase, which is written as
\begin{equation}
  \Pw{}(\ap,\wp) \Definition \dexp{\wp(\sqrt{1-\ap^2}-1)}.
\end{equation}
\citet[][chapter 4]{thompson86} shows that this additional phase can be
neglected only if\footnote{In contrast to the convention used in this
  paper, the \dmax{} and \dfield{} unit is meter instead of unit of
  \wavelength{} in the second form of the criterion, in order to
  explicitely show the dependence on the wavelength.}
\begin{equation}
  \frac{\pi}{4}\frac{\Afield^2}{\Asynth} \ll 1 \quad \mbox{or} \quad
  \pi\frac{\wavelength\dmax}{\dfield^2} \ll 1.
\end{equation}
The first form of the criterion indicates that the approximation gets worse
at high spatial dynamic range (\ie{}, $\Afield/\Asynth \ll 1$) while the
second form indicates that the approximation gets worse at long
wavelengths.

\subsection{$uvw$-unfaceting}

For stop-and-go mosaicking, it is usual to delay-track at the center of the
primary beam for each pointing/field of the mosaic. This phase center is
also the natural center of projection of each pointing/field.  Stop-and-go
mosaicking thus naturally paves the celestial sphere with as many tangent
planes as there are pointings/fields; \ie{}, this observing scheme is
somehow enforcing a $uvw$-faceting scheme. In the framework of on-the-fly
observations with ALMA, \citet{daddario00} indicate that the phase center
will be modified between each on-the-fly scan while it will stay constant
during each on-the-fly scan. This is a compromise between loss of coherence
and technical possibilities of the phase-locked loop. Using this
hypothesis, the maximum sky area covered by the on-the-fly scan must take
into account the maximum tolerable $w$-axis distortion.

The easiest way to deal with such data is to image each pointing/field
around its phase center and then to reproject this image onto the mosaic
tangent plane as displayed on Fig.~5 of~\citet{sault96b}. These authors
point out that this scheme implies a typical $w$-axis distortion $\epsilon$
less than
\begin{equation}
  \epsilon \le (1-\cos\Aalias)\,\sin\Acenter \sim \frac{1}{2} \,\Acenter\,\Aalias^2,
\end{equation}
where \Acenter{} is the angle from the pointing/field center and \Aalias{}
is the anti-aliasing scale defined in Sect.~\ref{sec:resampling}. In
particular, $\epsilon$ is 0 at the phase center of each pointing/field. In
other words, this scheme limits the magnitude of the $w$-axis distortion to
its magnitude on a size equal to the anti-aliasing scale (\ie{}, a few time
the primary beamwidth) instead of a size equal to the total mosaic field of
view.  This scheme thus solves the projection effect as long as the
$w$-axis distortion is negligible at sizes smaller than or equal to the
anti-aliasing scale.  A natural name for this processing scheme is
$uvw$-unfaceting because it is the combination of a faceting observing mode
(\ie{}, regular change of phase center) and a linear transform of the $uv$
coordinates to derive a single sine projection for the whole field of view.

\citet{sault96b} also demonstrate that the reprojection may be done much
more easily and quickly in the $uvw$ space before imaging the visibilities
because it is then just a simple transformation of the $uv$ coordinates,
followed by a multiplication of the visibilities by a phase term. Finally,
\citet{sault96b} note that it is the linear character of this $uv$
coordinate transform which preserves the measurement
equation~\ref{eq:measurement:otf}. As the change of coordinates happens
before any other processing, it also conserves all the equations derived in
the previous sections to implement the wide-field synthesis.

\section{On-the-fly observing mode and effective primary beam}
\label{sec:otf}

Usual interferometric observing modes (including stop-and-go mosaicking)
implies that the interferometer antennas observe a fixed point of the sky
during the integration time. Conversely, the on-the-fly observing mode
implies that the antennas slew on the sky during the integration time. This
implies that the measurement equation~\ref{eq:measurement:otf} must be
written as~\citep{holdaway94,rodriguez09}:\\
\V{}(\upave,\asave) =
\begin{equation}
  \Tave{\tone}{\ttwo}{\!\bracket{\!\sint{\ap}{\!\!\!\!\!\B{}\!\cbrace{\ap-\as(t)}\,\I{}(\ap)\,\dexp{\ap\up(t)}}}\!\!},
  \label{eq:measurement:otf:2}
\end{equation}
where \dt{} is the integration time and \upave{} and \asave{} are the mean
spatial frequency and direction cosine, defined as
\begin{equation}
  \upave \Definition \Tave{\tone}{\ttwo}{\!\!\!\!\!\up(t)} \quad \mbox{and} \quad
  \asave \Definition \Tave{\tone}{\ttwo}{\!\!\!\!\!\as(t)}.
\end{equation}
In this section, we analyze the consequences of the antenna slewing on the
accuracy of the wide-field synthesis.

\subsection{Time averaging}

\TabSymbolOTF{} %

In all interferometric observing modes, it is usual to adjust the
integration time so that $\up(t)$ can be approximated as \upave{}. To do
this, it is enough to ensure that $\up(t)$ always varies less than the $uv$
distance associated with tolerable aliasing (\dalias{}, see
Sect.~\ref{sec:resampling}) during the integration time (\dt{})
\begin{equation}
  \dt \ll \frac{\dalias}{\dmax\,\wearth} \quad \mbox{or} \quad
  \frac{dt}{1\,\mbox{sec}} \ll\frac{6900}{\Aalias/\Asynth},
  \label{eq:cond}
\end{equation}
where \dmax{} is the maximum baseline length, \wearth{} is the angular
velocity of a spatial frequency due to the Earth rotation ($7.27 \times
10^{-5}$ rad$\,$s$^{-1}$), \Aalias{} and \Asynth{} are respectively the
minimum field of view giving a tolerable aliasing and the synthesized beam
angular values.

\subsection{Effective primary beam}

Assuming that condition~\ref{eq:cond} is ensured, we can write
Eq.~\ref{eq:measurement:otf:2} with the same form as
Eq.~\ref{eq:measurement:otf} by the introduction of an effective primary
beam (\Beff{}); \ie{},
\begin{equation}
  \V{}(\upave,\asave) = \sint{\ap}{\Beff(\ap-\asave)\,\I{}(\ap)\,\dexp{\ap\upave}},
\end{equation}
\begin{equation}
  \mbox{where }
  \Beff(\ap-\asave) \Definition \Tave{\tone}{\ttwo}{\B{}\!\cbrace{\ap-\as(t)}}.
\end{equation}
Using the following change of variables
\begin{equation}
  \ab \Definition \as(t)-\asave, \quad
  d\ab = \frac{d\as(t)}{dt}dt \quad \mbox{or} \quad
  dt = \frac{d\ab}{\vslew(\ab)},
\end{equation}
we derive
\begin{equation}
  \Beff(\ap-\asave) = \sint{\ab}{\B{}\cbrace{(\ap-\asave)-\ab}\,\A{}(\ab)}
\end{equation}
\begin{equation}
  \mbox{with} \quad
  \A{}(\ab) \Definition \frac{1}{\vslew(\ab)\,\dt}\,\boxcar\paren{\frac{\ab}{\daslew}}
\end{equation}
\begin{equation}
  \mbox{and} \quad
   \daslew \Definition \int_{\tone}^{\ttwo} \vslew(t) \,\df{t}.
\end{equation}
In these equations, $\vslew(\ab)$ is the slew angular velocity of the
telescope as a function of the sky position, \daslew{} is the angular
distance covered during \dt{}, \A{} is an apodizing function, and
$\boxcar(\ab)$ is the usual rectangle function, which reproduces the finite
character of the time integration.

\subsection{Interpretation} 

The form of the measurement equation is conserved when averaging the
visibility function over a finite integration time, as long as the true
primary beam is replaced by an effective primary beam, which is the
convolution of the true primary beam by an apodizing function. To go
further, it is important to return to the two dimensional case. Indeed, the
convolution must be done along the slewing direction, resulting in an
effective primary beam elongated in a particular direction.

In principle, the equations derived in Sect.~\ref{sec:ekers-rots} can be
accommodated just by replacing the true primary beam by its effective
associate. In practice, the probability to take into account the effective
primary beam is low because its shape varies with time. Indeed, it is often
assumed that the sky is slewed along a straight line at constant angular
velocity. Even in this simplest case, it is advisable to slew along at
least two perpendicular directions to average systematic errors, implying
two different effective primary beams. However, practical reasons may/will
lead to complex scanning patterns: 1) The limitation of the acceleration
when trying to image a square region leads to spiral or Lissajous scanning
patterns; 2) The probable absence of derotators in future multi-beam
receivers (B.~Lazareff, private communication) implies the need to take
into account the Earth rotation in the scanning patterns of the off-axis
pixels.

\subsection{Approximation accuracy} 

\FigEffBeam{} %

In the following, we thus ask what is the trade-off accuracy of using the
true primary beam instead of the effective primary beam. The first point to
mention is that using different scanning patterns somehow helps because the
averaging process then makes the bias less systematic. Following
\citet{holdaway94}, we quantify the accuracy lost in the Fourier plane.
Indeed, the Ekers \& Rots scheme tries to estimate missing sky brightness
Fourier components from their measurements apodized by the Fourier
transform of the primary beam. In the Fourier space, the above convolution
just translates into a product. The Fourier transform of the apodizing
function thus degrades the sensitivity of the measured visibility,
\V{}(\up,\as), to spatial frequencies at the edges of the interval
$[\up-\dprim,\up+\dprim]$. To guide us in our quantification of the
accuracy lost, we now explore the simplest case of linear scanning at
constant velocity, where $\vslew(\ab)$ is constant and $\daslew =
\vslew\,\dt$. The Fourier transform of the apodizing function is then a
sinc function:
\begin{equation}
  \Aft{}(\uf) = \sinc(\uf\,\daslew).
\end{equation}
The relative error implied by the use of the true primary beam instead of
the effective primary beam is then
\begin{equation}
  \frac{\Beffft{}(\uf)-\Bft{}(\uf)}{\Beffft{}(\uf)}
  = 1-\frac{1}{\Aft{}(\uf)}
  = 1-\frac{1}{\sinc(\uf\,\daslew)}.
\end{equation}
Fig.~\ref{fig:EffBeam} shows this relative error as a function of the
number of samples per primary beam FWHM in the image plane (\ie{},
$\Afwhm/\daslew$) for different $uv$ distances (in units of \dprim{}). We
see that we derive a 1\% accuracy at all \uf{} when we sample the image
plane at a rate of 5 dumps per primary beam. However getting a 0.1\%
accuracy needs quite high sampling rates (about 15). This must be compared
with the accuracy of knowledge of \Bft{}.

We note that if a better accuracy is needed than the one achievable with
the highest sampling rate, it is in theory possible to replace in the
correlator software the rectangle apodizing function by another function
which falls more smoothly. To avoid the loss of sensitivity inherent to the
use of such an apodizing function (by throwing away data at the edges of
the time interval of integration), would require, for instance, to
half-overlap the integration intervals. This would imply more book-keeping
in the correlator software and some noise correlation between the measured
visibilities.

\end{document}

